\documentclass[%
 reprint,
 amsmath,amssymb,
 aps,
prb,
floatfix,
]{revtex4-2}

\usepackage{graphicx}% Include figure files
\usepackage{dcolumn}% Align table columns on decimal point
\usepackage{bm}% bold math
\begin{document}

\preprint{APS/123-QED}

\title{Hematite Thin Films Grown on Z-Cut and Y-Cut Lithium Niobate Piezoelectric Substrates by Pulsed Laser Deposition}% Force line breaks with \\

\author{Maximilian Mihm}
    \email[Correspondence email address: ]{maximilian.mihm@physik.uni-augsburg.de}% Your name
    \affiliation{Institute of Physics, University of Augsburg, Universitätsstraße 1, 86159 Augsburg, Germany}%Lines break automatically or can be forced with \\
\author{Stephan Glamsch}%
\affiliation{Institute of Physics, University of Augsburg, Universitätsstraße 1, 86159 Augsburg, Germany}
\author{Christian Holzmann}%
\affiliation{Institute of Physics, University of Augsburg, Universitätsstraße 1, 86159 Augsburg, Germany}
\author{Matthias Küß}%
\affiliation{Institute of Physics, University of Augsburg, Universitätsstraße 1, 86159 Augsburg, Germany}
\author{Helmut Karl}%
\affiliation{Institute of Physics, University of Augsburg, Universitätsstraße 1, 86159 Augsburg, Germany}
\author{Manfred Albrecht}%
\affiliation{Institute of Physics, University of Augsburg, Universitätsstraße 1, 86159 Augsburg, Germany}

\date{\today}% It is always \today, today,
             %  but any date may be explicitly specified

\begin{abstract}
Altermagnets are a newly identified class of materials that combine advantageous characteristics of both ferro‑ and antiferromagnets, making them highly promising for spintronic applications. Hematite has recently been identified as an altermagnetic material and exhibits several noteworthy properties, including a high Néel temperature, a temperature dependent spin reorientation transition (SRT) at the Morin temperature ($T_\mathrm{M}$), and low magnetic damping. In this work, we demonstrate the epitaxial growth of hematite thin films on y- and z-cut lithium niobate (LiNbO$_3$) substrates using pulsed laser deposition (PLD). LiNbO$_3$ as piezoelectric substrate is of particular interest as it enables the efficient excitation of surface acoustic waves (SAWs) with interdigital transducers. 
The different substrate cuts allow for different orientations of the Néel vector.
Films grown on y-cut LiNbO$_3$ are single-crystalline and single-phase, while those deposited on z-cut LiNbO$_3$ exhibit two distinct in-plane (ip) domains rotated $60\,$° relative to each other. 
On both substrates, the hematite thin films exhibit a temperature dependent SRT which allows the antiferromagnetic Néel vector to be controlled. This study paves the way for the development of high-quality piezoelectric/altermagnetic hyprids for magnonics and spintronics.

\end{abstract}

%\keywords{Suggested keywords}%Use showkeys class option if keyword
                              %display desired
\maketitle

%\tableofcontents

\section{Introduction}
Hematite ($\alpha$-Fe$_2$O$_3$) is a well-studied antiferromagnetic insulator with a Néel temperature of approximately $950\,$K \cite{Morin1950, Dannegger2023}. It crystallizes in the $R\bar{3}c$ space group with the lattice parameters $a = 0.5035\,$nm and $c = 1.3747\,$nm \cite{Maslen1994}.  
Its low damping coefficient, comparable to that of yttrium iron garnet \cite{Hamdi2023, Lebrun2020}, makes it a prime candidate for antiferromagnetic magnonics \cite{Hamdi2023, Lebrun2020, ElKanj2023}. 
Due to the Dzyaloshinskii-Moriya interaction, hematite is a canted antiferromagnet at room temperature, resulting in a small net magnetic moment \cite{Dzyaloshinsky1958, Moriya1960}. Around $260\,$K bulk $\alpha$-Fe$_2$O$_3$ undergoes a SRT, known as the Morin transition \cite{Morin1950}. 
Above the Morin temperature, the spin axis lies within the hexagonal $ab$‑plane, whereas below $T_\mathrm{M}$ the spins align collinearly and antiparallel to the $c$-axis. 
The main reasons for the SRT are the magnetic-dipole anisotropy and the single-ion anisotropy \cite{Artman1965, Levinson1969}.  
Modifying either contribution allows $T_\mathrm{M}$ to be shifted to higher or lower temperatures. Moreover, the Morin temperature can be influenced through elemental doping or strain, as demonstrated for bulk materials and nanoparticles \cite{Hayashi2021, Curry1965, Nakamura1964, Kren1965, Besser1967, Vandenberghe1986, Popov2025, Krehula2017}. Similar effects of doping on the SRT have also been observed in thin films \cite{Tanaka2024, Shimomura2015, Nozaki2019, Ellis2017}. Moreover, in thin film systems, strain can be tuned through the use of different substrates \cite{Serrano2018, Park2013, TodaCasaban2025}, and $T_{M}$ can even be shifted above room temperature. Additionally, several studies have shown that varying the hematite film thickness enables further control of $T_M$ \cite{Tanaka2024, Shimomura2015, Park2013, Liu2025}. Furthermore, it has been reported that the Morin transition does not occur when the out-of-plane (oop) lattice strain is compressive \cite{Park2013, Liu2025, Kan2022}.
Hematite has been theoretically identified as an altermagnet \cite{Smejkal2022, Smejkal2023, Hoyer2025}, a prediction that was recently experimentally confirmed by X-ray photoemission microscopy and anomalous Hall transport measurements \cite{Galindez‐Ruales2025}. 
While the effects of static strain in antiferromagnets are well established \cite{Song2018, Yan2019}, and its influence on altermagnets is becoming increasingly explored \cite{Aoyama2024, Karetta2025, Zhang2025, Chakraborty2024}, investigations of dynamic strain in antiferromagnets and altermagnets remain comparatively scarce. This is despite the fact that striking phenomena, such as the acoustic spin‑splitter effect, have been theoretically predicted in altermagnets \cite{Gunnink2026}. However, realizing such experiments requires an altermagnet/piezoelectric hybrid structure, which is challenging to fabricate. 
In this study, we demonstrate the epitaxial growth of altermagnetic hematite thin films on piezoelectric z-cut LiNbO$_3$(0001) and y-cut LiNbO$_3$(1$\bar{1}$00) substrates using PLD. LiNbO$_3$ is a well-established material for SAW devices \cite{Morgan2010}.
Epitaxial growth is facilitated by the fact that LiNbO$_3$ crystallizes in the same space group ($R\bar{3}c$) as hematite and exhibits a small lattice mismatch of $2.2\,$\% along the $a$-direction, and $0.83\,$\% along the $c$-direction ($a = 0.5148\,$nm and $c = 1.3861\,$nm \cite{Abrahams1966}). For comparison, the lattice mismatch between Al$_2$O$_3$ and hematite along the $a$-direction is $-5.6\,$\% and along the $c$-direction is $-5.8\,$\%, even though Al$_2$O$_3$ is a widely used substrate for the epitaxial growth of hematite thin films \cite{ Tanaka2024, Shimomura2015, Nozaki2019,  Serrano2018, Park2013, Liu2025, Kan2022, Scheufele2023, Qiu2023}. 
We systematically investigate the film growth over a wide range of deposition temperatures and O$_2$ pressures. Furthermore, we analyze the magnetic properties of the grown films, including the Morin transition and the spin configuration in dependence of the film orientation.
Although previous studies have reported single-crystalline hematite thin films on LiNbO$_3$(0001) substrates prepared by either mist chemical vapor deposition \cite{Shimazoe2020} or  by magnetron reactive radio frequency sputtering \cite{Luzanov2022}, the magnetic properties of such altermagnetic/piezoelectric hybrids have not been investigated.

\section{Experimental}

Thin film deposition was performed using a PLD setup. The laser used was a KrF excimer laser (Coherent ComPEX 205F) with a wavelength of 248 nm, applying laser pulses with a repetition rate of $3\,$Hz and a pulse duration of $30\,$ns. For all films, the laser energy was $550\,$mJ and the fluence was set to $2.3\,$J$\,$cm$^{-2}$ \cite{Jung2022}. All films were deposited using a polycrystalline $\alpha$-Fe$_2$O$_3$ target. The  target was prepared from $\alpha$-Fe$_2$O$_3$ powder (99.9$\,\%$, ChemPUR) pressed into a pellet and sintered for 15 hours at 1000$\,$°C in air. To investigate the influence of the substrate temperature and oxygen partial pressure on the growth individually, a temperature series and an O$_2$ pressure series were carried out. For the temperature series, the substrate temperature was varied in 50$\,$°C steps between 425 and 625$\,$°C while the oxygen partial pressure was kept at 2\texttimes$10^{-4}\,$mbar. For the pressure series, the substrate temperature was $575\,$°C and the oxygen partial pressure was varied between 2\texttimes$10^{-5}$ and 2\texttimes$10^{-2}\,$mbar. During each deposition run, a y-cut and a z-cut LiNbO$_3$ substrate were coated simultaneously to ensure the same deposition conditions.   \\ XRD measurements were performed using a Rigaku Smartlab $9\,$kW system with a rotating copper anode (Cu$_{K_{\alpha}}$ with a wavelength of $0.1541\,$nm) to determine the structure and phase formation of the iron oxide thin films. The film thickness was determined by XRR. The XRR curves were fitted the SmartLab Studio II software from Rigaku. AFM images were recorded using a Dimension Icon AFM instrument from Bruker. EBSD measurements were performed using a Zeiss Merlin scanning electron microscope (SEM) equipped with a Symmetry S2 detector from Oxford Instruments.\\$M$ vs $T$ measurements were performed using a superconducting quantum interference device-vibrating sample magnetometer (SQUID-VSM, MPMS3, Quantum Design).

\section{Results and Discussion}
\subsection{Structural Characterization}

\begin{figure}[!ht]
	\centering
	\includegraphics{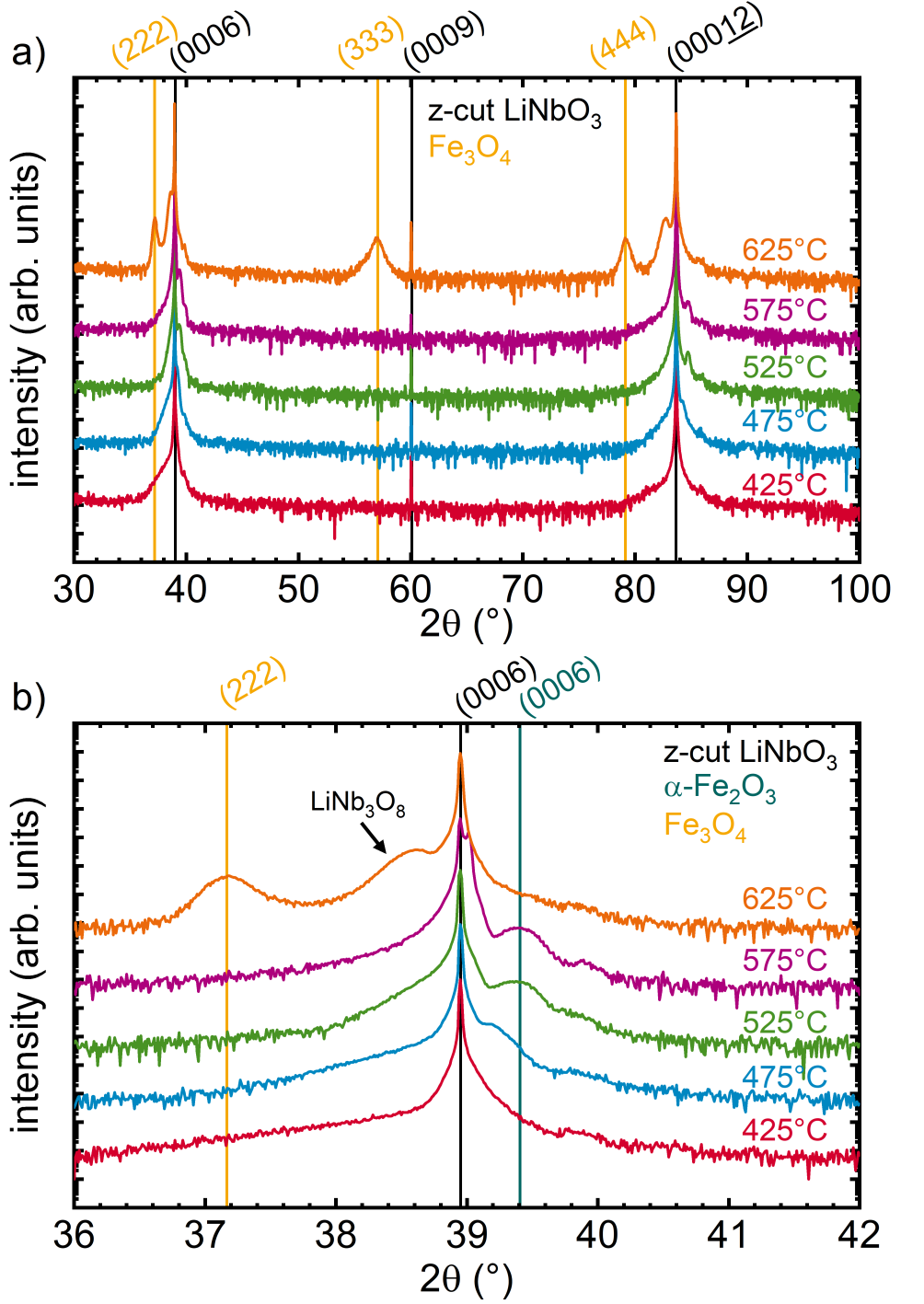}
	\caption{XRD pattern of hematite films deposited at different temperatures on z-cut LiNbO$_3$ with an oxygen partial pressure of 2\texttimes$10^{-4}\,$mbar. a) Overview and b) enlargement around the (0006) substrate peak.}
	\label{fig:1}
\end{figure}

\begin{figure}[!ht]
	\centering
	\includegraphics{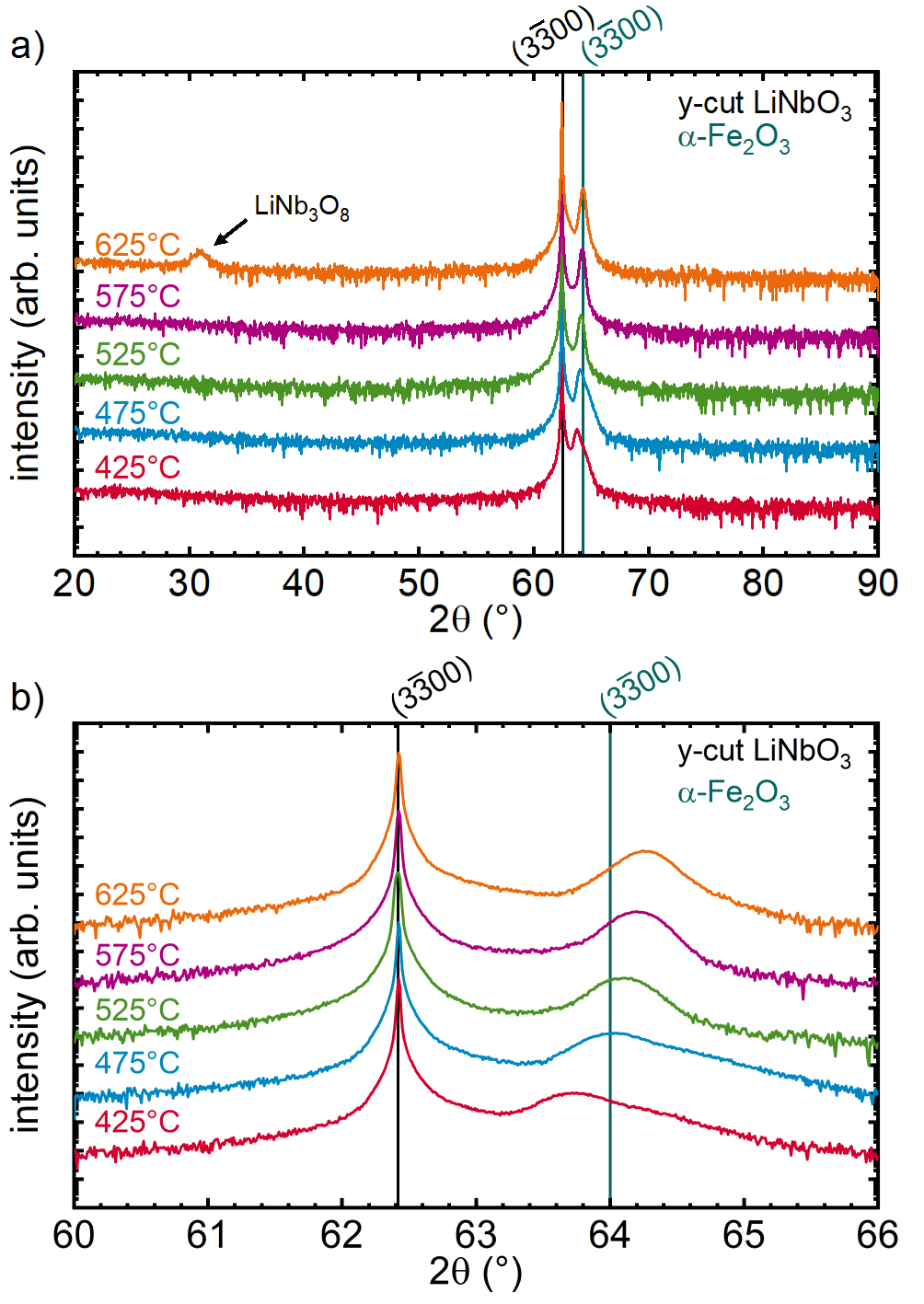}
	\caption{XRD pattern of hematite films deposited at different temperatures on y-cut LiNbO$_3$ with an oxygen partial pressure of 2\texttimes$10^{-4}\,$mbar. a) Overview and b) enlargement of the (3$\bar{3}$00) peak.}
	\label{fig:2}
\end{figure}

\begin{figure}[!hb]
	\centering
	\includegraphics{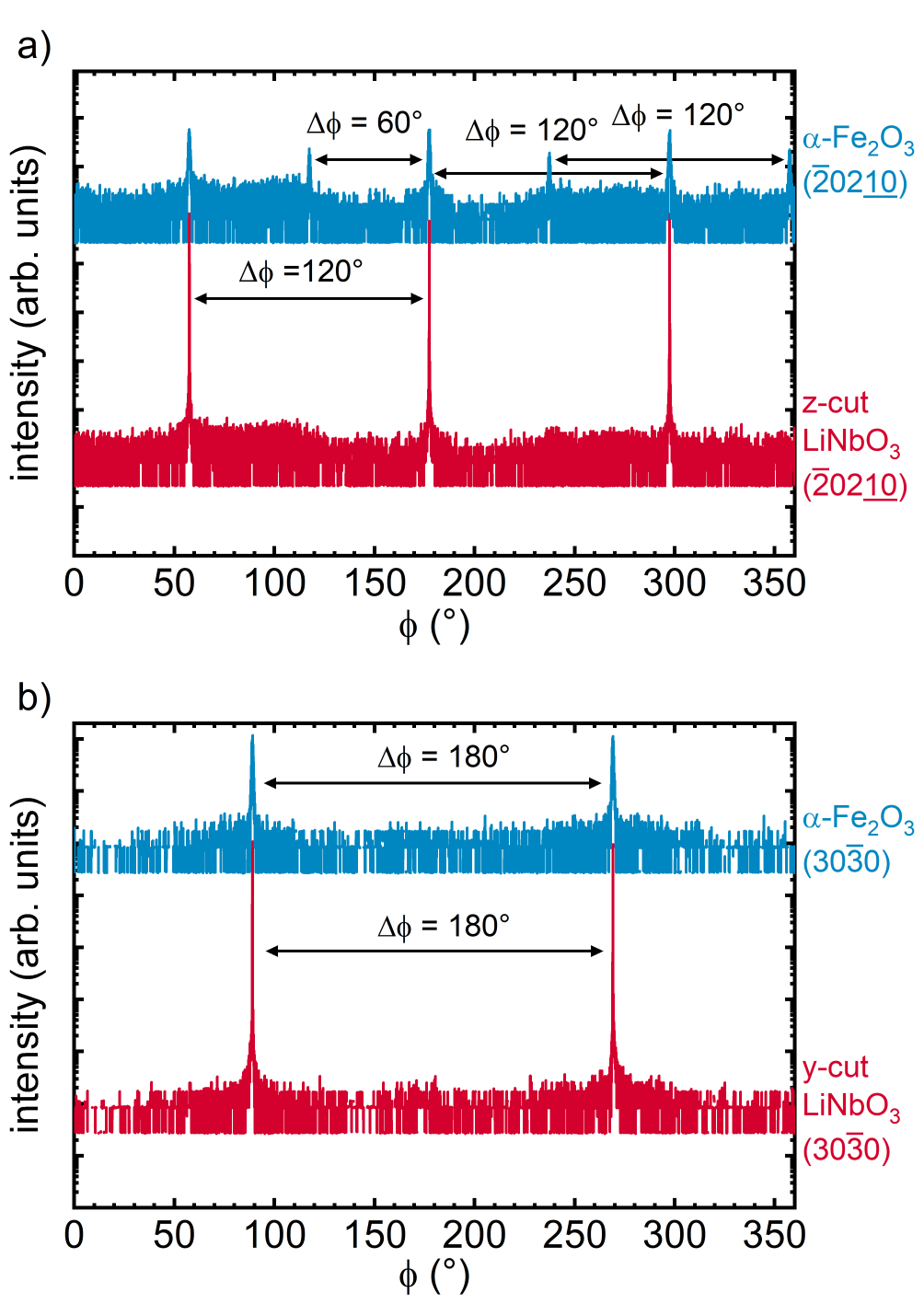}
	\caption{$\phi$-scans for hematite grown on a) z-cut and b) y-cut LiNbO$_3$ at $575\,$°C, 2\texttimes$10^{-4}\,$mbar. For the z-cut LiNbO$_3$, three substrate reflections, corresponding to the ($\bar{2}$02\underline{10}) reflection are observed. For hematite six peaks are observed, spaced $60\,$° apart. For the thin film on y-cut LiNbO$_3$ $\phi$-scans around the (30$\bar{3}$0) reflection were performed. Both, film and substrate, showing only two reflections that are perfectly aligned.}
	\label{fig:3}
\end{figure}

 \textbf{Figure} \ref{fig:1} shows $2\theta/\omega $ X-ray diffraction (XRD) patterns of hematite thin films grown on z-cut LiNbO$_3$ at different temperatures (temperature series). At a substrate temperature of $425\,$°C the film peak is weak and very close to the substrate reflection, appearing as a shoulder on the right side, see Figure \ref{fig:1}b). By increasing the substrate temperature by $50\,$°C the peaks are getting more pronounced, which belong to the (0006) and (000\underline{12}) reflections of hematite. With further increase of the temperature the peaks become even more pronounced as the crystallinity increases. The hematite (0006) peak shifts to higher 2$\theta$ angles with increasing temperature, which means the film gets more compressed in the oop direction. However, at  a substrate temperature of $625\,$°C no hematite reflections are observed anymore. Instead, new peaks around $37\,$°, $57\,$°, and $79\,$° appear belonging to magnetite (Fe$_3$O$_4$), which is consistent with results reported for Al$_2$O$_3$(0001) \cite{Bejjit2022, Tiwari2007}. Additionally, the peak around $38.6\,$° belongs to LiNb$_3$O$_8$, which forms at higher temperatures in reduced atmosphere \cite{Armenise1983,Jackel1981,McCoy1994, Namkoong2005}. 
\textbf{Figure} \ref{fig:2}a) shows the XRD pattern of hematite films grown on y-cut LiNbO$_3$. These samples show a similar temperature dependent behavior as the hematite thin films on z-cut LiNbO$_3$. With increasing deposition temperature the (3$\bar{3}$00) film peak shifts to higher $2\theta$ angles (see Figure \ref{fig:2}b)). At a substrate temperature of $625\,$°C LiNb$_3$O$_8$ also forms on the y-cut LiNbO$_3$. The difference of the y-cut and z-cut LiNbO$_3$ is the $c$-axis orientation. For the z-cut LiNbO$_3$, the $c$-axis points perpendicular to the substrate surface, while for the y-cut LiNbO$_3$, the $c$-axis lies ip. Because of the shifting 2$\theta$ angle, the $c$ lattice parameter decreases with increasing temperature for films grown on z-cut LiNbO$_3$, while for hematite thin films on y-cut LiNbO$_3$ the $a$ lattice parameter decreases with increasing temperature.
 \\To investigate the influence of the oxygen partial pressure on the film growth, a pressure series was performed. The substrate temperature was kept at $575\,$°C and the O$_2$ pressure was varied between 2\texttimes$10^{-5}\,$mbar and 2\texttimes$10^{-2}\,$mbar. The XRD patterns for these films grown on z-cut and y-cut LiNbO$_3$ substrates can be found in the supplementary material \textbf{Figures} S1 and S2, respectively. On the z-cut LiNbO$_3$, hematite could be stabilized for oxygen partial pressures higher than 2\texttimes$10^{-5}\,$mbar. At an oxygen partial pressure of 2\texttimes$10^{-5}\,$mbar Fe$_3$O$_4$, and additionally the LiNb$_3$O$_8$ phase formed. In addition, hematite thin films on y-cut LiNbO$_3$ can be stabilized throughout the studied pressure range. Consequently, hematite has a larger growth window on y-cut than on z-cut LiNbO$_3$.
 \\All grown films have thicknesses between 39 and $62\,$nm as extracted from X-ray reflectrometry (XRR) data, see \textbf{Table} S1 (temperature series) and S2 (pressure series) in the supplementary material. Due to the position of the samples on the sample holder, the thicknesses of the films can slightly vary although the films were deposited during the same run \cite{Mihm2025}. An exemplary XRR fit is shown in \textbf{Figure} S3. \\
 To study the ip relationship between the films and the different substrates, $\phi$-scans were performed \cite{Mihm2025, Holzmann2025a}. For hematite grown at $575\,$°C with an oxygen partial pressure of 2\texttimes$10^{-4}\,$mbar on y-cut and z-cut LiNbO$_3$ the (30$\bar{3}$0) and ($\bar{2}$02\underline{10}) reflections were used, with resulting $\phi$-scans shown in \textbf{Figure} \ref{fig:3}a) (z-cut) and \ref{fig:3}b) (y-cut). Hematite films grown on y-cut substrates exhibit ip aligned epitaxy, while films grown on z-cut LiNbO$_3$ show two ip domains, which are rotated by $60\,$°. This is in contrast to hematite grown on Al$_2$O$_3$(0001), where the films showed perfect ip epitaxial growth \cite{Serrano2018, TodaCasaban2025, Kan2022}. In addition, hematite grown on SrTiO$_3$(111) also showed two different ip domains but rotated by $\pm30\,$° relative to the substrate \cite{Serrano2018, TodaCasaban2025}. 
 \begin{figure*}[!htb]
	\centering
	\includegraphics{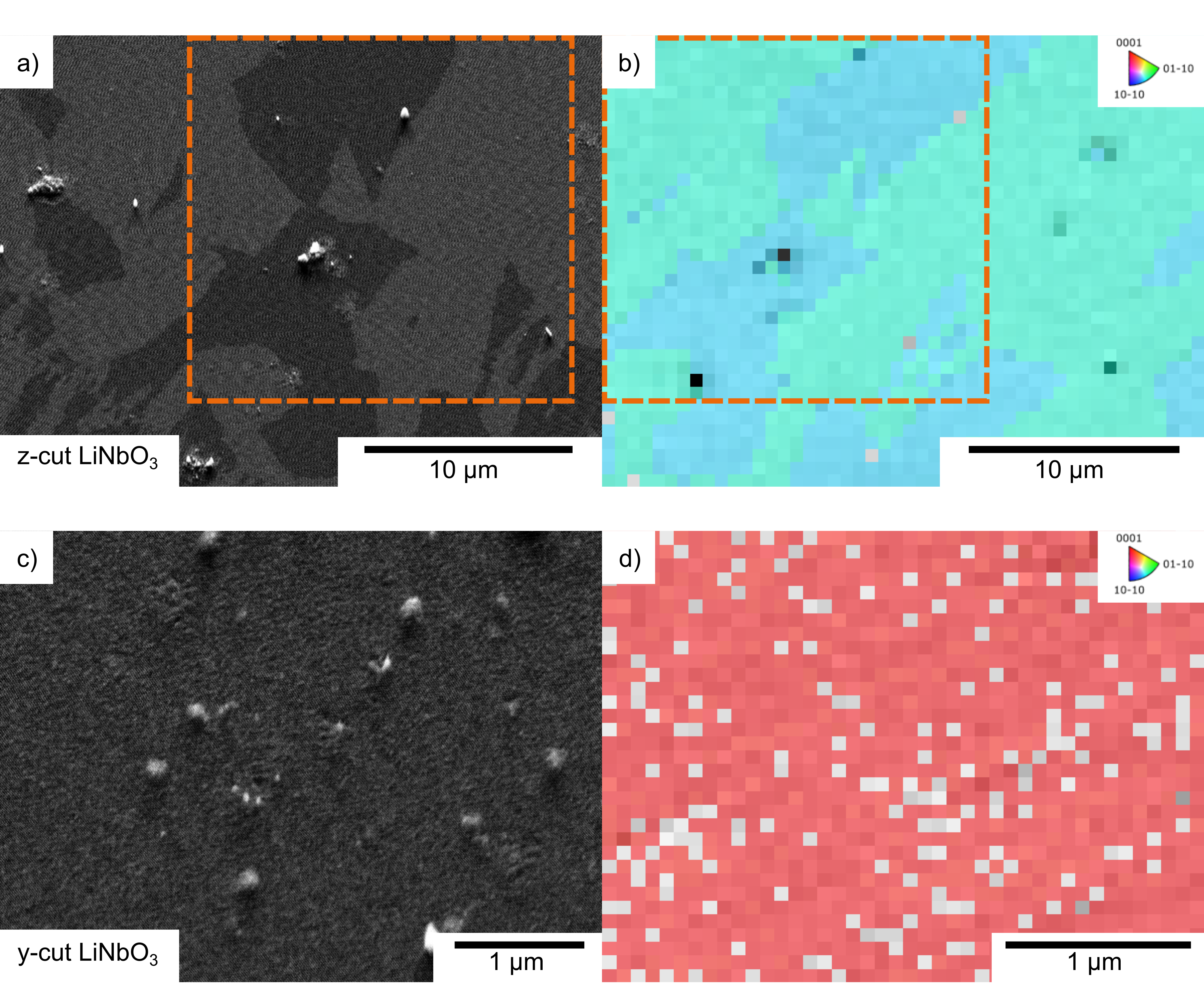}
	\caption{Images of the microstructure of hematite thin films, grown at $475\,$°C and an O$_2$ pressure of 2\texttimes$10^{-4}\,$mbar, on a) z-cut and c) y-cut LiNbO$_3$ using a FSD. The orange rectangles in a) and b) are guidance for the eye to see the correlation between FSD and EBSD image. Panels b) and d) show the corresponding IPF coloring along an ip direction for hematite on z-cut and y-cut LiNbO$_3$, respectively. Gray pixels correspond to areas where no hematite was identified either no Kikuchi patterns were detected or they could not be assigned to hematite. }
	\label{fig:4}
\end{figure*}
 \\To investigate these domains in more detail, electron backscatter diffraction (EBSD) measurements were performed, on hematite films that were grown at $475\,$°C (oxygen partial pressure of 2\texttimes$10^{-4}\,$mbar) on z-cut and y-cut LiNbO$_3$. \textbf{Figure} \ref{fig:4}a) and c) show the corresponding forward scatter detector (FSD) images. The FSD images reveal the microstructure of the grown films. In Figure \ref{fig:4}b) and d) the corresponding inverse pole figure (IPF) coloring maps are displayed. Please note that the gray pixels correspond to areas where no hematite was identified either no Kikuchi patterns were detected or they could not be assigned to hematite. For hematite on z-cut LiNbO$_3$ two domains with different contrasts are visible, which have a size of up to $10\,$µm. The two different contrasts in Figure \ref{fig:4}a) fit perfectly to the contrast in the IPF image in Figure \ref{fig:4}b). Blue displays the $<\bar{1}$$\bar{1}$20$>$ and turquoise presents the $<1$$\bar{2}$10$>$ ip direction. In contrast, for the film grown on y-cut LiNbO$_3$ only one domain is visible in the FSD image, and thus only one color is observed in the corresponding IPF map (Figure \ref{fig:4}d), which fits to the observations of the $\phi$-scans. 
 \\ In a further study, the morphology of the iron oxide films were recorded by atomic force microscopy (AFM). The images are shown in the supplementary material, \textbf{Figures} S4 and S5 (temperature series), and \textbf{Figures} S6 and S7 (pressure series)). The root mean square (RMS) roughness for all grown films is summarized in the supplementary material in \textbf{Table} S3 (temperature series) and \textbf{Table} S4 (pressure series). The RMS values for all films are between 0.32 and $1.80\,$nm, which is comparable to other oxides grown by PLD \cite{Serrano2018, TodaCasaban2025, Holzmann2025, Mihm2025, Holzmann2025a, Holzmann2022}. It should be noted that films grown on z-cut LiNbO$_3$ have a lower RMS roughness than those grown on y-cut LiNbO$_3$. Films grown on z-cut LiNbO$_3$ with substrate temperatures lower than $500\,$°C or O$_2$ pressures between 2\texttimes$10^{-4}\,$mbar and 2\texttimes$10^{-3}\,$mbar have the smoothest surface with RMS roughness between $0.32\,$nm and $0.40\,$nm. For films grown on y-cut LiNbO$_3$, the smoothest surface with an RMS roughness of $0.49\,$nm was achieved using a substrate temperature of $575\,$°C and an oxygen partial pressure of 2\texttimes$10^{-2}\,$mbar.

\subsection{Magnetic Properties}
\begin{figure*}[!htp]
	\centering
	\includegraphics{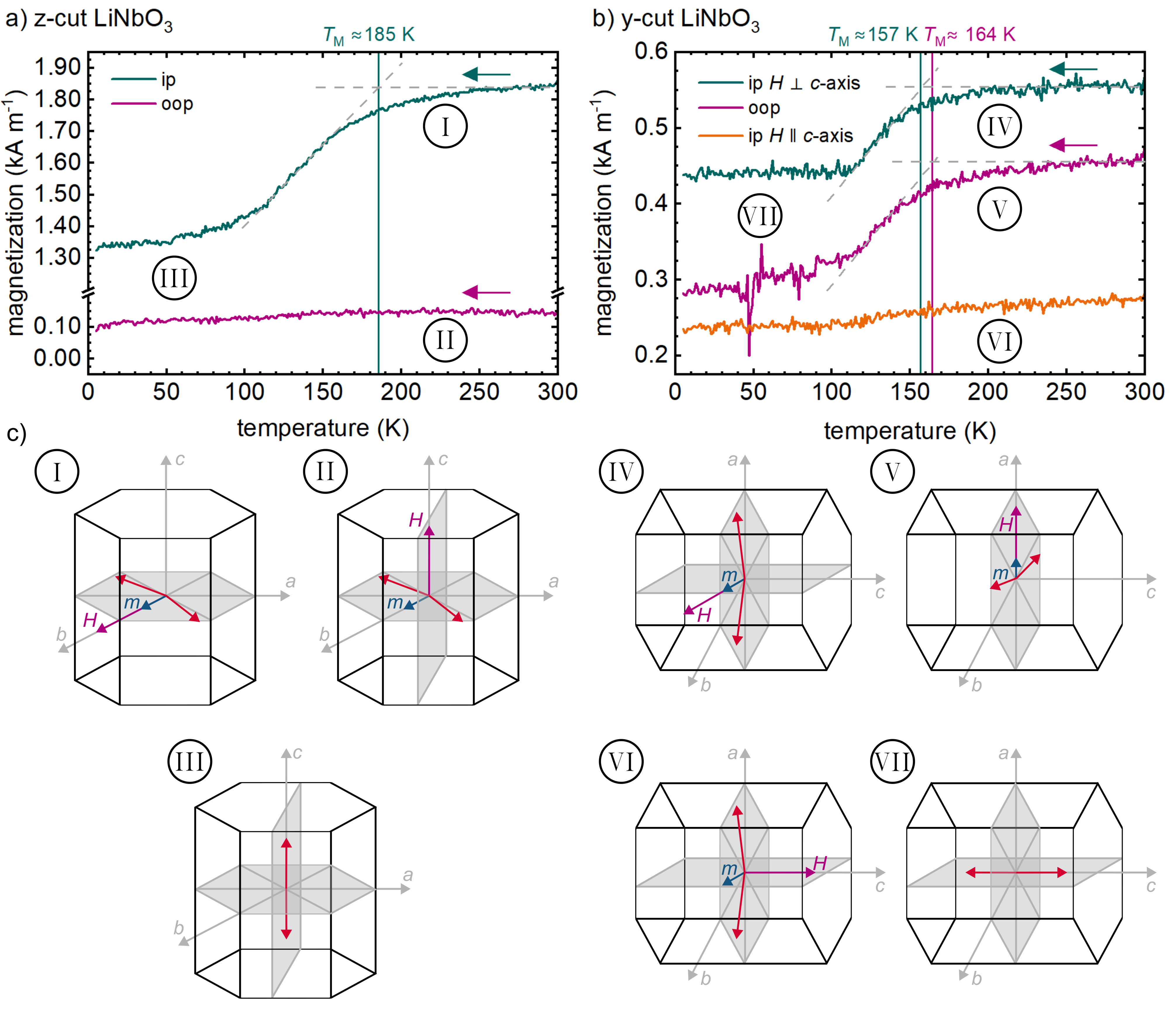}
	\caption{$M$ vs $T$ measurements along the ip and oop directions of hematite films grown at $575\,$°C with an oxygen partial pressure of 2\texttimes$10^{-3}\,$mbar on a) z-cut and b) y-cut LiNbO$_3$ substrates. Hematite films on y-cut and z-cut LiNbO$_3$ were 62 and $61\,$nm thick, respectively. Samples were magnetized before the measurement and the magnetization versus temperature was measured with no guiding field during cooling. c) Schematic spin alignment of hematite grown on z-cut and y-cut LiNbO$_3$ substrates above and below the Morin temperature. The arrows in the hexagonal unit cell indicating directions of the spins for the different substrate cuts and measurements configurations above and below $T_\mathrm{M}$. $H$ is the applied saturation magnetic field before the measurement, aligned either ip or oop, and $m$ is the small net moment arising from spin canting.}
	\label{fig:5}
\end{figure*}
To investigate the influence of the different structural features of hematite thin films on the Morin transition and its spin configuration, magnetization vs temperature ($M$ vs $T$) measurements were performed. Prior to the measurement, the films were magnetized in a $7\,$T field either in the ip or oop direction and then cooled from 300 to $5\,$K at rate of $2\,$K$\,\mathrm{min}^{-1}$ without an applied field. During the cooling process, the magnetization was recorded in the direction of the initial magnetic field. Please note that no background correction was performed. \textbf{Figure} \ref{fig:5}a) shows the $M$ vs $T$ curves for a $61\,$nm thick hematite film grown at $575\,$°C with an O$_2$ pressure of 2\texttimes$10^{-3}\,$mbar on a z-cut LiNbO$_3$ substrate. During the ip measurement, a drop in magnetization between $185\,$K and $100\,$K is observed, which corresponds to the Morin transition. We define the $T_\mathrm{M}$ as the intersection of two straight lines (gray in Figure \ref{fig:5}a)): one line with constant magnetization, the other with decreasing magnetization. This yields a $T_\mathrm{M}$ value of approximately $185\,$K. This value is about $80\,$K lower than the value reported for bulk material \cite{Morin1950}. Similar values have been determined for hematite thin films on Al$_2$O$_3$(0001) \cite{Tanaka2024, Liu2025, Gota2001, Galindez‐Ruales2025}. The lower Morin temperature compared to bulk material could result from the low film thickness \cite{Tanaka2024, Shimomura2015,Park2013, Liu2025, Gota2001, Scheufele2023} or the stress within the film \cite{Park2013, Liu2025}. Due to the lattice mismatch between hematite and LiNbO$_3$, the film is strained compressively along the ip direction, leading to a tensile strain along the oop direction. Park et al. reported that an ip compressive strain can influence the Morin temperature \cite{Park2013}. This lattice expansion in the $c$-direction results in a change of the relative Fe$^{3+}$ ions positions, which affects the dipolar anisotropy and can pin the magnetic moments in the basal plane. This effect can suppress or even prevent the occurrence of a Morin transition \cite{Gota2001}. Furthermore, the transition region is relatively broad, and it appears, that the SRT is not fully completed. We attribute this to the magnetic contribution of the substrate, as the substrates magnetic signal decreases at lower temperatures (see \textbf{Figure} S8). In addition, we performed $M$ vs $T$ measurement along the oop $c$-direction. No SRT was observed here. It should be noted that the difference in magnetization values obtained in the ip and oop direction is mainly due to the measurement geometry. Furthermore, the two different ip domains for hematite films grown on z-cut LiNbO$_3$ substrates should have no influence on the measurement since the spins lie in the $ab$-plane with a six fold anisotropy, where the moments align with the $a$-axes \cite{Hamdi2023, Besser1967}. Due to the magnetization before the measurement all spins are aligned along the field direction. 
\\Based on these observations, the underlying spin configurations can now be discussed. Above the Morin temperature, the canted antiferromagnetic spins are aligned within the hexagonal $ab$-plane. By applying a magnetic field in this plane, the resulting net magnetic moment of the canted spins will follow the field direction, in this case along the $b$-axis in the plane, as displayed in \ref{fig:5}c) (I). In contrast, below the Morin transition, the antiferromagnetic spins are collinearly aligned along the oop $c$-axis (Figure \ref{fig:5}c) (III)), so the resulting net moment will vanish. $M$ vs $T$ measurements along the oop $c$-axis yield no magnetic signal either below or above the Morin temperature, as expected from the underlying spin configuration (see Figure \ref{fig:5}c)(II, III)). It should be noted that above the Morin temperature, the net moment remains in the $ab$-plane even after the applied oop magnetic field is switched off. 
\\Figure \ref{fig:5}b) shows the $M$ vs $T$ curves for a $62\,$nm thick hematite film on y-cut LiNbO$_3$, which was grown in the same deposition run to ensure the same growth conditions. The situation here is similar, but different in that the $ab$-plane is now orientated parallel to the oop direction. Therefore, the Morin transition is observed for the ip measurement, when a field is applied perpendicular to the $c$-axis, and for an oop measurement geometry. For a field applied parallel to the $c$-axis, no SRT is detected. Above the Morin temperature, the applied field can align the resulting net magnetic moment in both geometries, as displayed in Figure \ref{fig:5}c) (IV, V). Suturin et al. observed the rotation of the Néel vector of thin hematite films, measured by X-ray magnetic linear dichroism (XMLD), by applying an external field in different directions \cite{Suturin2021}. The Néel vector stands perpendicular to the small ferromagnetic moment. Therefore, these results fit our observations because the small net magnetic moment can be aligned ip or oop, above $T_\mathrm{M}$. For an initial field applied parallel to the $c$-axis, the spins are likely to remain in the $ab$-plane and therefore the small magnetic moment as well (see Figure \ref{fig:5}c) (VII)). Below the Morin temperature, due to the collinear alignment of the antiferromagnetic spins along the $c$-axis, which now points into the film plane, a vanishing net moment is again observed (see Figure \ref{fig:5}c) (VII)).  Furthermore, the SRT starts at approximately $160\,$K, which is about $25\,$K lower than that of the film grown on the z-cut substrate. This could be due to the different stress levels in these films. Furthermore, for comparison, we also measured $M$ vs $T$ curves on blank LiNbO$_3$ substrates, see Figure S8, which reveal some small features at $50\,$K or $100\,$K attributable to internal strain effects \cite{FernandezRuiz2005}. These can sometimes also be seen in our thin film samples (see Figure \ref{fig:5}b)) and originate from the substrate.
\section{Conclusion}
In summary, we have epitaxially grown hematite thin films on piezoelectric LiNbO$_3$ substrates, with thicknesses between $39\,$nm and $62\,$nm using PLD. The hematite films grown on y-cut LiNbO$_3$ are single-crystal and single-phase, exhibiting epitaxial growth with aligned ip crystallographic axes. The films grown on z-cut substrates are also single-phase but feature two ip domains which are rotated by 60$\,$° relative to each other. Furthermore, the RMS roughness of all grown films is well below $2\,$nm. On both substrate cuts, the films exhibit a SRT. For films grown on y-cut LiNbO$_3$, the Morin temperature is slightly lower ($T_\mathrm{M_{oop}}\sim 160\,$K) than for films grown on z-cut substrates ($T_\mathrm{M_{ip}}\sim 185\,$K). The canted antiferromagnetic spins are oriented in the ab‑plane above $T_\mathrm{M}$ and become collinear along the $c$‑axis below $T_\mathrm{M}$. The $c$‑axis alignment, and thus the overall spin orientation, can be controlled through the choice of LiNbO$_3$ substrate cut. For hematite thin films grown on z-cut LiNbO$_3$, the canted antiferromagnetic spins lie in the $ab$-plane above the Morin temperature and below they are orientated collinear along the $c$-axis. In contrast, for films grown on y-cut LiNbO$_3$ the $ab$-plane is now oriented parallel to the oop direction. Thus, above the SRT the spins will align in the $ab$-plane, so that the net magnetic moment will points either ip or oop, depending on the direction of the initially applied magnetic field. Below the SRT, the spins are again oriented collinearly along the $c$-axis, which is now pointing along the film plane. This study paves the way for the development of high‑qualtiy piezoelectric/altermagnetic hybrids thin‑film devices for magnonics and spintronics, including the possibility of controlling altermagnetic properties using dynamically generated strain from SAWs.
\section*{Acknowledgment}
This project was partly funded by the Deutsche Forschungsgemeinschaft (DFG, German Research Foundation) project numbers 318592081, 470034807, and 540566574.
\section*{Data Availability}
The data are available from the authors upon reasonable
request.
\bibliography{hematite}% Produces the bibliography via BibTeX.

@Article{Lebrun2020,
  author    = {Lebrun, R. and Ross, A. and Gomonay, O. and Baltz, V. and Ebels, U. and Barra, A.-L. and Qaiumzadeh, A. and Brataas, A. and Sinova, J. and Kläui, M.},
  journal   = {Nature Communications},
  title     = {{Long-distance spin-transport across the Morin phase transition up to room temperature in ultra-low damping single crystals of the antiferromagnet $\alpha$-Fe$_{2}$O$_{3}$}},
  year      = {2020},
  issn      = {2041-1723},
  month     = dec,
  number    = {1},
  pages     = {6332},
  volume    = {11},
  doi       = {10.1038/s41467-020-20155-7},
  publisher = {Springer Science and Business Media LLC},
}

@Article{ElKanj2023,
  author    = {El Kanj, Aya and Gomonay, Olena and Boventer, Isabella and Bortolotti, Paolo and Cros, Vincent and Anane, Abdelmadjid and Lebrun, Romain},
  journal   = {Science Advances},
  title     = {{Antiferromagnetic magnon spintronic based on nonreciprocal and nondegenerated ultra-fast spin-waves in the canted antiferromagnet $\alpha$-Fe$_{2}$O$_{3}$}},
  year      = {2023},
  issn      = {2375-2548},
  month     = aug,
  number    = {32},
  pages     = {eadh1601},
  volume    = {9},
  doi       = {10.1126/sciadv.adh1601},
  publisher = {American Association for the Advancement of Science (AAAS)},
}

@Article{Shimazoe2020,
  author    = {Shimazoe, Kazuki and Nishinaka, Hiroyuki and Arata, Yuta and Tahara, Daisuke and Yoshimoto, Masahiro},
  journal   = {AIP Advances},
  title     = {{Phase control of $\alpha$- and $\kappa$-Ga$_{2}$O$_{3}$ epitaxial growth on LiNbO$_{3}$ and LiTaO$_{3}$ substrates using $\alpha$-Fe$_{2}$O$_{3}$ buffer layers}},
  year      = {2020},
  issn      = {2158-3226},
  month     = may,
  number    = {5},
  pages     = {055310},
  volume    = {10},
  doi       = {10.1063/5.0006137},
  publisher = {AIP Publishing},
}

@Article{Luzanov2022,
  author    = {Luzanov, V. A.},
  journal   = {Journal of Communications Technology and Electronics},
  title     = {{Growth of Epitaxial Fe$_{2}$O$_{3}$ Films on Lithium Niobate Substrates}},
  year      = {2022},
  issn      = {1555-6557},
  month     = mar,
  number    = {3},
  pages     = {296--297},
  volume    = {67},
  doi       = {10.1134/s106422692203010x},
  publisher = {Pleiades Publishing Ltd},
}

@Article{Galindez‐Ruales2025,
  author    = {Galindez‐Ruales, Edgar and Gonzalez‐Hernandez, Rafael and Schmitt, Christin and Das, Shubhankar and Fuhrmann, Felix and Ross, Andrew and Golias, Evangelos and Akashdeep, Akashdeep and Lünenbürger, Laura and Baek, Eunchong and Yang, Wanting and Šmejkal, Libor and Krishna, Venkata and Jaeschke‐Ubiergo, Rodrigo and Sinova, Jairo and Rothschild, Avner and You, Chun‐Yeol and Jakob, Gerhard and Kläui, Mathias},
  journal   = {Advanced Materials},
  title     = {{Revealing the Altermagnetism in Hematite via XMCD Imaging and Anomalous Hall Electrical Transport}},
  year      = {2025},
  issn      = {1521-4095},
  month     = jul,
  pages     = {e05019},
  volume    = {37},
  doi       = {10.1002/adma.202505019},
  publisher = {Wiley},
}

@Article{Hamdi2023,
  author    = {Hamdi, Mohammad and Posva, Ferdinand and Grundler, Dirk},
  journal   = {Physical Review Materials},
  title     = {{Spin wave dispersion of ultra-low damping hematite ($\alpha$-Fe$_{2}$O$_{3}$) at GHz frequencies}},
  year      = {2023},
  issn      = {2475-9953},
  month     = may,
  number    = {5},
  pages     = {054407},
  volume    = {7},
  doi       = {10.1103/physrevmaterials.7.054407},
  publisher = {American Physical Society (APS)},
}

@Article{Hoyer2025,
  author    = {Hoyer, Rhea and Stavropoulos, P. Peter and Razpopov, Aleksandar and Valentí, Roser and Šmejkal, Libor and Mook, Alexander},
  journal   = {Physical Review B},
  title     = {{Altermagnetic splitting of magnons in hematite $\alpha$-Fe$_{2}$O$_{3}$}},
  year      = {2025},
  issn      = {2469-9969},
  month     = aug,
  number    = {6},
  pages     = {064425},
  volume    = {112},
  doi       = {10.1103/fgc1-5blp},
  publisher = {American Physical Society (APS)},
}

@Article{Smejkal2022,
  author    = {Šmejkal, Libor and Sinova, Jairo and Jungwirth, Tomas},
  journal   = {Physical Review X},
  title     = {{Beyond Conventional Ferromagnetism and Antiferromagnetism: A Phase with Nonrelativistic Spin and Crystal Rotation Symmetry}},
  year      = {2022},
  issn      = {2160-3308},
  month     = sep,
  number    = {3},
  pages     = {031042},
  volume    = {12},
  doi       = {10.1103/physrevx.12.031042},
  publisher = {American Physical Society (APS)},
}

@Article{Bejjit2022,
  author    = {Bejjit, C-E and Rogé, V. and Cachoncinlle, C. and Hebert, C. and Perrière, J. and Briand, E. and Millon, E.},
  journal   = {Thin Solid Films},
  title     = {{Iron oxide thin films grown on (00l) sapphire substrate by pulsed-laser deposition}},
  year      = {2022},
  issn      = {0040-6090},
  month     = mar,
  pages     = {139101},
  volume    = {745},
  doi       = {10.1016/j.tsf.2022.139101},
  publisher = {Elsevier BV},
}

@Article{Qiu2023,
  author    = {Qiu, Hongsong and Seifert, Tom S. and Huang, Lin and Zhou, Yongjian and Kašpar, Zdeněk and Zhang, Caihong and Wu, Jingbo and Fan, Kebin and Zhang, Qi and Wu, Di and Kampfrath, Tobias and Song, Cheng and Jin, Biaobing and Chen, Jian and Wu, Peiheng},
  journal   = {Advanced Science},
  title     = {{Terahertz Spin Current Dynamics in Antiferromagnetic Hematite}},
  year      = {2023},
  issn      = {2198-3844},
  month     = apr,
  number    = {18},
  pages     = {2300512},
  volume    = {10},
  doi       = {10.1002/advs.202300512},
  publisher = {Wiley},
}

@Article{Tiwari2007,
  author    = {Tiwari, Shailja and Prakash, Ram and Choudhary, R J and Phase, D M},
  journal   = {Journal of Physics D: Applied Physics},
  title     = {{Oriented growth of Fe$_{3}$O$_{4}$thin film on crystalline and amorphous substrates by pulsed laser deposition}},
  year      = {2007},
  issn      = {1361-6463},
  month     = aug,
  number    = {16},
  pages     = {4943--4947},
  volume    = {40},
  doi       = {10.1088/0022-3727/40/16/028},
  publisher = {IOP Publishing},
}

@Article{Scheufele2023,
  author    = {Scheufele, M. and Gückelhorn, J. and Opel, M. and Kamra, A. and Huebl, H. and Gross, R. and Geprägs, S. and Althammer, M.},
  journal   = {APL Materials},
  title     = {{Impact of growth conditions on magnetic anisotropy and magnon Hanle effect in $\alpha$-Fe$_{2}$O$_{3}$}},
  year      = {2023},
  issn      = {2166-532X},
  month     = sep,
  number    = {9},
  pages     = {091115},
  volume    = {11},
  doi       = {10.1063/5.0160304},
  publisher = {AIP Publishing},
}

@Article{Liu2025,
  author    = {Liu, Haoyu and Zhang, Hantao and Keagy, Josiah and Gao, Qinwu and Li, Letian and Li, Junxue and Cheng, Ran and Shi, Jing},
  journal   = {Physical Review Materials},
  title     = {{Anisotropic field suppression of Morin transition temperature in epitaxially grown hematite thin films}},
  year      = {2025},
  issn      = {2475-9953},
  month     = mar,
  number    = {3},
  pages     = {034410},
  volume    = {9},
  doi       = {10.1103/physrevmaterials.9.034410},
  publisher = {American Physical Society (APS)},
}

@Article{Suturin2021,
  author    = {Suturin, Sergey M. and Korovin, Alexander M. and Gastev, Sergey V. and Dvortsova, Polina A. and Volkov, Mikhail P. and Valvidares, Manuel and Sokolov, Nikolai S.},
  journal   = {Physical Review Materials},
  title     = {{X-ray magnetic linear dichroism study of field-manipulated canted antiferromagnetism in epitaxial $\alpha$-Fe$_{2}$O$_{3}$ films}},
  year      = {2021},
  issn      = {2475-9953},
  month     = apr,
  number    = {4},
  pages     = {044408},
  volume    = {5},
  doi       = {10.1103/physrevmaterials.5.044408},
  publisher = {American Physical Society (APS)},
}

@Article{Gota2001,
  author    = {Gota, Susana and Gautier-Soyer, Martine and Sacchi, Maurizio},
  journal   = {Physical Review B},
  title     = {{Magnetic properties of Fe$_{2}$O$_{3}$(0001) thin layers studied by soft x-ray linear dichroism}},
  year      = {2001},
  issn      = {1095-3795},
  month     = nov,
  number    = {22},
  pages     = {224407},
  volume    = {64},
  doi       = {10.1103/physrevb.64.224407},
  publisher = {American Physical Society (APS)},
}

@Article{Hayashi2021,
  author    = {Hayashi, Kensuke and Yamada, Keisuke and Shima, Mutsuhiro and Ohya, Yutaka and Ono, Teruo and Moriyama, Takahiro},
  journal   = {Applied Physics Letters},
  title     = {{Control of antiferromagnetic resonance and the Morin temperature in cation doped $\alpha$ -Fe$_{2-x}$M$_{x}$O$_{3}$ (M = Al, Ru, Rh, and In)}},
  year      = {2021},
  issn      = {1077-3118},
  month     = jul,
  number    = {3},
  pages     = {032408},
  volume    = {119},
  doi       = {10.1063/5.0053586},
  publisher = {AIP Publishing},
}

@Article{Kan2022,
  author    = {Kan, Daisuke and Moriyama, Takahiro and Aso, Ryotaro and Horai, Shinji and Shimakawa, Yuichi},
  journal   = {Applied Physics Letters},
  title     = {{Triaxial magnetic anisotropy and Morin transition in $\alpha$-Fe$_{2}$O$_{3}$ epitaxial films characterized by spin Hall magnetoresistance}},
  year      = {2022},
  issn      = {1077-3118},
  month     = mar,
  number    = {11},
  pages     = {112403},
  volume    = {120},
  doi       = {10.1063/5.0087643},
  publisher = {AIP Publishing},
}

@Article{Nozaki2019,
  author    = {Nozaki, Tomohiro and Pati, Satya Prakash and Shiokawa, Yohei and Suzuki, Motohiro and Ina, Toshiaki and Mibu, Ko and Al-Mahdawi, Muftah and Ye, Shujun and Sahashi, Masashi},
  journal   = {Journal of Applied Physics},
  title     = {{Identifying valency and occupation sites of Ir dopants in antiferromagnetic $\alpha$-Fe$_{2}$O$_{3}$ thin films with X-ray absorption fine structure and Mössbauer spectroscopy}},
  year      = {2019},
  issn      = {1089-7550},
  month     = mar,
  number    = {11},
  pages     = {113903},
  volume    = {125},
  doi       = {10.1063/1.5080483},
  publisher = {AIP Publishing},
}

@Article{Dannegger2023,
  author    = {Dannegger, Tobias and Deák, András and Rózsa, Levente and Galindez-Ruales, E. and Das, Shubhankar and Baek, Eunchong and Kläui, Mathias and Szunyogh, László and Nowak, Ulrich},
  journal   = {Physical Review B},
  title     = {{Magnetic properties of hematite revealed by an \textit{ab initio} parameterized spin model}},
  year      = {2023},
  issn      = {2469-9969},
  month     = may,
  number    = {18},
  pages     = {184426},
  volume    = {107},
  doi       = {10.1103/physrevb.107.184426},
  publisher = {American Physical Society (APS)},
}

@Article{Smejkal2023,
  author    = {Šmejkal, Libor and Marmodoro, Alberto and Ahn, Kyo-Hoon and González-Hernández, Rafael and Turek, Ilja and Mankovsky, Sergiy and Ebert, Hubert and D’Souza, Sunil W. and Šipr, Ondřej and Sinova, Jairo and Jungwirth, Tomáš},
  journal   = {Physical Review Letters},
  title     = {{Chiral Magnons in Altermagnetic RuO$_{2}$}},
  year      = {2023},
  issn      = {1079-7114},
  month     = dec,
  number    = {25},
  pages     = {256703},
  volume    = {131},
  doi       = {10.1103/physrevlett.131.256703},
  publisher = {American Physical Society (APS)},
}

@Article{Morin1950,
  author    = {Morin, F. J.},
  journal   = {Physical Review},
  title     = {{Magnetic Susceptibility of $\alpha$Fe$_{2}$O$_{3}$ and $\alpha$Fe$_{2}$O$_{3}$ with Added Titanium}},
  year      = {1950},
  issn      = {0031-899X},
  month     = jun,
  number    = {6},
  pages     = {819--820},
  volume    = {78},
  doi       = {10.1103/physrev.78.819.2},
  publisher = {American Physical Society (APS)},
}

@Article{Mihm2025,
  author    = {Mihm, Maximilian and Holzmann, Christian and Seyd, Johannes and Ullrich, Aladin and Karl, Helmut and Albrecht, Manfred},
  journal   = {Thin Solid Films},
  title     = {{Phase Control of Single-Crystalline Cobalt Oxide Thin Films Grown by Pulsed Laser Deposition}},
  year      = {2025},
  issn      = {0040-6090},
  month     = oct,
  pages     = {140776},
  volume    = {827},
  doi       = {10.1016/j.tsf.2025.140776},
  publisher = {Elsevier BV},
}

@Article{Abrahams1966,
  author    = {Abrahams, S.C. and Hamilton, W.C. and Reddy, J.M.},
  journal   = {Journal of Physics and Chemistry of Solids},
  title     = {{Ferroelectric lithium niobate. 4. Single crystal neutron diffraction study at 24°C}},
  year      = {1966},
  issn      = {0022-3697},
  month     = jun,
  number    = {6–7},
  pages     = {1013--1018},
  volume    = {27},
  doi       = {10.1016/0022-3697(66)90073-4},
  publisher = {Elsevier BV},
}

@Article{Maslen1994,
  author    = {Maslen, E. N. and Streltsov, V. A. and Streltsova, N. R. and Ishizawa, N.},
  journal   = {Acta Crystallographica Section B Structural Science},
  title     = {{Synchrotron X-ray study of the electron density in $\alpha$-Fe$_{2}$O$_{3}$}},
  year      = {1994},
  issn      = {0108-7681},
  month     = aug,
  number    = {4},
  pages     = {435--441},
  volume    = {50},
  doi       = {10.1107/s0108768194002284},
  publisher = {International Union of Crystallography (IUCr)},
}

@Article{Curry1965,
  author    = {Curry, N. A. and Johnston, G. B. and Besser, P. J. and Morrish, A. H.},
  journal   = {Philosophical Magazine},
  title     = {{Neutron diffraction measurements on pure and doped synthetic hematite crystals}},
  year      = {1965},
  issn      = {0031-8086},
  month     = aug,
  number    = {116},
  pages     = {221--228},
  volume    = {12},
  doi       = {10.1080/14786436508218865},
  publisher = {Informa UK Limited},
}

@Article{Levinson1969,
  author    = {Levinson, Lionel M. and Luban, Marshall and Shtrikman, S.},
  journal   = {Physical Review},
  title     = {{Microscopic Model for Reorientation of the Easy Axis of Magnetization}},
  year      = {1969},
  issn      = {0031-899X},
  month     = nov,
  number    = {2},
  pages     = {715--722},
  volume    = {187},
  doi       = {10.1103/physrev.187.715},
  publisher = {American Physical Society (APS)},
}

@Article{Artman1965,
  author    = {Artman, J. O. and Murphy, J. C. and Foner, S.},
  journal   = {Physical Review},
  title     = {{Magnetic Anisotropy in Antiferromagnetic Corundum-Type Sesquioxides}},
  year      = {1965},
  issn      = {0031-899X},
  month     = may,
  number    = {3A},
  pages     = {A912--A917},
  volume    = {138},
  doi       = {10.1103/physrev.138.a912},
  publisher = {American Physical Society (APS)},
}

@Article{Nakamura1964,
  author    = {Nakamura, T. and Seinjo, T. and Endoh, Y. and Yamamoto, N. and Shiga, M. and Nakamura, Y.},
  journal   = {Physics Letters},
  title     = {{Fe$^{57}$ Mössbauer effect in ultra fine particles of $\alpha$-Fe$_{2}$O$_{3}$}},
  year      = {1964},
  issn      = {0031-9163},
  month     = oct,
  number    = {3},
  pages     = {178--179},
  volume    = {12},
  doi       = {10.1016/0031-9163(64)91054-6},
  publisher = {Elsevier BV},
}

@Article{Kren1965,
  author    = {Krén, E. and Szabó, P. and Konczos, G.},
  journal   = {Physics Letters},
  title     = {{Neutron diffraction studies on the (1-x) Fe$_{2}$O$_{3}$ - xRh$_{2}$O$_{3}$ system}},
  year      = {1965},
  issn      = {0031-9163},
  month     = oct,
  number    = {2},
  pages     = {103--104},
  volume    = {19},
  doi       = {10.1016/0031-9163(65)90731-6},
  publisher = {Elsevier BV},
}

@Article{Besser1967,
  author    = {Besser, P. J. and Morrish, A. H. and Searle, C. W.},
  journal   = {Physical Review},
  title     = {{Magnetocrystalline Anisotropy of Pure and Doped Hematite}},
  year      = {1967},
  issn      = {0031-899X},
  month     = jan,
  number    = {2},
  pages     = {632--640},
  volume    = {153},
  doi       = {10.1103/physrev.153.632},
  publisher = {American Physical Society (APS)},
}

@Article{Vandenberghe1986,
  author    = {Vandenberghe, R.E. and Verbeeck, A.E. and De Grave, E.},
  journal   = {Journal of Magnetism and Magnetic Materials},
  title     = {{On the Morin transition in Mn-substituted hematite}},
  year      = {1986},
  issn      = {0304-8853},
  month     = feb,
  number    = {Part 2},
  pages     = {898--900},
  volume    = {54–57},
  doi       = {10.1016/0304-8853(86)90304-5},
  publisher = {Elsevier BV},
}

@Article{Popov2025,
  author    = {Popov, Nina and Marijan, Sara and Pavić, Luka and Miljanić, Snežana and Zadro, Krešo and Kratofil Krehula, Ljerka and Homonnay, Zoltán and Kuzmann, Ernő and Kubuki, Shiro and Ibrahim, Ahmed and Krehula, Stjepko},
  journal   = {Journal of Alloys and Compounds},
  title     = {{Influence of Al$^{3+}$ ions on the direct hydrothermal formation and properties of hematite ($\alpha$-Fe$_{2}$O$_{3}$) nanorods}},
  year      = {2025},
  issn      = {0925-8388},
  month     = mar,
  pages     = {179223},
  volume    = {1018},
  doi       = {10.1016/j.jallcom.2025.179223},
  publisher = {Elsevier BV},
}

@Article{Krehula2017,
  author    = {Krehula, Stjepko and Ristić, Mira and Reissner, Michael and Kubuki, Shiro and Musić, Svetozar},
  journal   = {Journal of Alloys and Compounds},
  title     = {{Synthesis and properties of indium-doped hematite}},
  year      = {2017},
  issn      = {0925-8388},
  month     = feb,
  pages     = {1900--1907},
  volume    = {695},
  doi       = {10.1016/j.jallcom.2016.11.022},
  publisher = {Elsevier BV},
}

@Article{Armenise1983,
  author    = {Armenise, M. N. and Canali, C. and De Sario, M. and Carnera, A. and Mazzoldi, P. and Celotti, G.},
  journal   = {Journal of Applied Physics},
  title     = {{Characterization of TiO$_{2}$, LiNb$_{3}$O8, and (Ti$_{0.65}$Nb$_{0.35}$)O$_{2}$ compound growth observed during Ti:LiNbO$_{3}$ optical waveguide fabrication}},
  year      = {1983},
  issn      = {1089-7550},
  month     = nov,
  number    = {11},
  pages     = {6223--6231},
  volume    = {54},
  doi       = {10.1063/1.331939},
  publisher = {AIP Publishing},
}

@Article{Jackel1981,
  author    = {Jackel, J. L. and Ramaswamy, V. and Lyman, S. P.},
  journal   = {Applied Physics Letters},
  title     = {{Elimination of out-diffused surface guiding in titanium-diffused LiNbO$_{3}$}},
  year      = {1981},
  issn      = {1077-3118},
  month     = apr,
  number    = {7},
  pages     = {509--511},
  volume    = {38},
  doi       = {10.1063/1.92433},
  publisher = {AIP Publishing},
}

@Article{McCoy1994,
  author    = {McCoy, M.A. and Dregia, S.A. and Lee, W.E.},
  journal   = {Journal of Materials Research},
  title     = {{Crystallography of surface nucleation and epitaxial growth of lithium triniobate on congruent lithium niobate}},
  year      = {1994},
  issn      = {2044-5326},
  month     = aug,
  number    = {8},
  pages     = {2029--2039},
  volume    = {9},
  doi       = {10.1557/jmr.1994.2029},
  publisher = {Springer Science and Business Media LLC},
}

@Article{Namkoong2005,
  author    = {Namkoong, Gon and Lee, Kyoung-Keun and Madison, Shannon M. and Henderson, Walter and Ralph, Stephen E. and Doolittle, W. Alan},
  journal   = {Applied Physics Letters},
  title     = {{III-nitride integration on ferroelectric materials of lithium niobate by molecular beam epitaxy}},
  year      = {2005},
  issn      = {1077-3118},
  month     = oct,
  number    = {17},
  pages     = {171107},
  volume    = {87},
  doi       = {10.1063/1.2084340},
  publisher = {AIP Publishing},
}

@Article{FernandezRuiz2005,
  author    = {Fernández-Ruiz, R. and Martín y Marero, D. and Bermúdez, V.},
  journal   = {Physical Review B},
  title     = {{Anomalous structural feature of LiNbO$_{3}$ observed using neutron diffraction}},
  year      = {2005},
  issn      = {1550-235X},
  month     = nov,
  number    = {18},
  pages     = {184108},
  volume    = {72},
  doi       = {10.1103/physrevb.72.184108},
  publisher = {American Physical Society (APS)},
}

@Article{Ellis2017,
  author    = {Ellis, David S. and Weschke, Eugen and Kay, Asaf and Grave, Daniel A. and Malviya, Kirtiman Deo and Mor, Hadar and de Groot, Frank M. F. and Dotan, Hen and Rothschild, Avner},
  journal   = {Physical Review B},
  title     = {{Magnetic states at the surface of $\alpha$-Fe$_{2}$O$_{3}$ thin films doped with Ti, Zn, or Sn}},
  year      = {2017},
  issn      = {2469-9969},
  month     = sep,
  number    = {9},
  pages     = {094426},
  volume    = {96},
  doi       = {10.1103/physrevb.96.094426},
  publisher = {American Physical Society (APS)},
}

@Article{Holzmann2025a,
  author    = {Holzmann, Christian and Glamsch, Stephan and Stein, David and Mihm, Maximilian and Ullrich, Aladin and Schlitz, Richard and Lammel, Michaela and Boneberg, Johannes and Albrecht, Manfred},
  journal   = {Physical Review Materials},
  title     = {{Inverse garnet/Pt heterostructures by lateral crystallization}},
  year      = {2025},
  issn      = {2475-9953},
  month     = nov,
  number    = {11},
  volume    = {9},
  doi       = {10.1103/hhk6-qg6l},
  publisher = {American Physical Society (APS)},
}

@Book{Morgan2010,
  author    = {Morgan, David P.},
  publisher = {Elsevier Science \& Technology},
  title     = {{Surface Acoustic Wave Filters}},
  year      = {2010},
  address   = {San Diego},
  edition   = {2nd ed.},
  isbn      = {9780080550138},
  series    = {Studies in Electrical and Electronic Engineering Ser},
  pagetotal = {1448},
  ppn_gvk   = {1658192966},
  subtitle  = {With Applications to Electronic Communications and Signal Processing},
}

@Article{Song2018,
  author    = {Song, Cheng and You, Yunfeng and Chen, Xianzhe and Zhou, Xiaofeng and Wang, Yuyan and Pan, Feng},
  journal   = {Nanotechnology},
  title     = {{How to manipulate magnetic states of antiferromagnets}},
  year      = {2018},
  issn      = {1361-6528},
  month     = mar,
  number    = {11},
  pages     = {112001},
  volume    = {29},
  doi       = {10.1088/1361-6528/aaa812},
  publisher = {IOP Publishing},
}

@Article{Yan2019,
  author    = {Yan, Han and Feng, Zexin and Shang, Shunli and Wang, Xiaoning and Hu, Zexiang and Wang, Jinhua and Zhu, Zengwei and Wang, Hui and Chen, Zuhuang and Hua, Hui and Lu, Wenkuo and Wang, Jingmin and Qin, Peixin and Guo, Huixin and Zhou, Xiaorong and Leng, Zhaoguogang and Liu, Zikui and Jiang, Chengbao and Coey, Michael and Liu, Zhiqi},
  journal   = {Nature Nanotechnology},
  title     = {{A piezoelectric, strain-controlled antiferromagnetic memory insensitive to magnetic fields}},
  year      = {2019},
  issn      = {1748-3395},
  month     = jan,
  number    = {2},
  pages     = {131--136},
  volume    = {14},
  doi       = {10.1038/s41565-018-0339-0},
  publisher = {Springer Science and Business Media LLC},
}

@Article{Aoyama2024,
  author    = {Aoyama, Takuya and Ohgushi, Kenya},
  journal   = {Physical Review Materials},
  title     = {{Piezomagnetic properties in altermagnetic MnTe}},
  year      = {2024},
  issn      = {2475-9953},
  month     = apr,
  number    = {4},
  pages     = {l041402},
  volume    = {8},
  doi       = {10.1103/physrevmaterials.8.l041402},
  publisher = {American Physical Society (APS)},
}

@Article{Karetta2025,
  author    = {Karetta, Bennet and Verbeek, Xanthe H. and Jaeschke-Ubiergo, Rodrigo and Šmejkal, Libor and Sinova, Jairo},
  journal   = {Physical Review B},
  title     = {{Strain-controlled \textit{g} - to \textit{d} -wave transition in altermagnetic CrSb}},
  year      = {2025},
  issn      = {2469-9969},
  month     = sep,
  number    = {9},
  volume    = {112},
  doi       = {10.1103/pbbr-hwz4},
  publisher = {American Physical Society (APS)},
}

@Article{Zhang2025,
  author    = {Zhang, Wancheng and Zheng, Mingkun and Liu, Yong and Zhang, Zhenhua and Xiong, Rui and Lu, Zhihong},
  journal   = {Physical Review B},
  title     = {{Strain-induced nonrelativistic altermagnetic spin splitting effect}},
  year      = {2025},
  issn      = {2469-9969},
  month     = jul,
  number    = {2},
  volume    = {112},
  doi       = {10.1103/8zlt-mlms},
  publisher = {American Physical Society (APS)},
}

@Article{Chakraborty2024,
  author    = {Chakraborty, Atasi and González Hernández, Rafael and Šmejkal, Libor and Sinova, Jairo},
  journal   = {Physical Review B},
  title     = {{Strain-induced phase transition from antiferromagnet to altermagnet}},
  year      = {2024},
  issn      = {2469-9969},
  month     = apr,
  number    = {14},
  pages     = {144421},
  volume    = {109},
  doi       = {10.1103/physrevb.109.144421},
  publisher = {American Physical Society (APS)},
}

@Article{Holzmann2025,
  author    = {Holzmann, Christian and Küß, Matthias and Glamsch, Stephan and Stein, David and Kunz, Yannik and Weiler, Mathias and Albrecht, Manfred},
  journal   = {ACS Applied Materials \& Interfaces},
  title     = {{Polycrystalline YIG Thin Films on a Piezoelectric Substrate for Magnetoacoustic Hybrid Devices}},
  year      = {2025},
  issn      = {1944-8252},
  month     = oct,
  pages     = {58550--58558},
  volume    = {17},
  doi       = {10.1021/acsami.5c14928},
  publisher = {American Chemical Society (ACS)},
}

@Article{TodaCasaban2025,
  author    = {Toda-Casaban, Meritxell and Balcells, Lluis and Mestres, Narcís and Pomar, Alberto and Chen, Hui and Garzón Manjón, Alba and Arbiol, Jordi and Martínez, Benjamín and Frontera, Carlos},
  journal   = {Acta Materialia},
  title     = {{Substrate-driven structural coherence in epitaxial hematite thin films for spintronics}},
  year      = {2025},
  issn      = {1359-6454},
  month     = dec,
  pages     = {121613},
  volume    = {301},
  doi       = {10.1016/j.actamat.2025.121613},
  publisher = {Elsevier BV},
}

@Article{Holzmann2022,
  author    = {Holzmann, Christian and Ullrich, Aladin and Ciubotariu, Oana-Tereza and Albrecht, Manfred},
  journal   = {ACS Applied Nano Materials},
  title     = {{Stress-Induced Magnetic Properties of Gadolinium Iron Garnet Nanoscale-Thin Films: Implications for Spintronic Devices}},
  year      = {2022},
  issn      = {2574-0970},
  month     = jan,
  number    = {1},
  pages     = {1023--1033},
  volume    = {5},
  doi       = {10.1021/acsanm.1c03687},
  publisher = {American Chemical Society (ACS)},
}

@Article{Jung2022,
  author    = {Jung, Florian and Delmdahl, Ralph and Heymann, Andreas and Fischer, Max and Karl, Helmut},
  journal   = {Applied Physics A},
  title     = {{Surface evolution of crystalline SrTiO$_{3}$, LaAlO$_{3}$ and Y$_{3}$Al$_{5}$O$_{12}$ targets during pulsed laser ablation}},
  year      = {2022},
  issn      = {1432-0630},
  month     = aug,
  number    = {9},
  pages     = {750},
  volume    = {128},
  doi       = {10.1007/s00339-022-05805-5},
  publisher = {Springer Science and Business Media LLC},
}

@Article{Park2013,
  author    = {Park, SeongHun and Jang, H. and Kim, J.-Y. and Park, B.-G. and Koo, T.-Y. and Park, J.-H.},
  journal   = {EPL (Europhysics Letters)},
  title     = {{Strain control of Morin temperature in epitaxial $\alpha$ -Fe$_{2}$O$_{3}$ (0001) film}},
  year      = {2013},
  issn      = {1286-4854},
  month     = jul,
  number    = {2},
  pages     = {27007},
  volume    = {103},
  doi       = {10.1209/0295-5075/103/27007},
  publisher = {IOP Publishing},
}

@Article{Serrano2018,
  author    = {Serrano, Aida and Rubio-Zuazo, Juan and López-Sánchez, Jesús and Arnay, Iciar and Salas-Colera, Eduardo and Castro, Germán R.},
  journal   = {The Journal of Physical Chemistry C},
  title     = {{Stabilization of Epitaxial $\alpha$-Fe$_{2}$O$_{3}$ Thin Films Grown by Pulsed Laser Deposition on Oxide Substrates}},
  year      = {2018},
  issn      = {1932-7455},
  month     = jun,
  number    = {28},
  pages     = {16042--16047},
  volume    = {122},
  doi       = {10.1021/acs.jpcc.8b02430},
  publisher = {American Chemical Society (ACS)},
}

@Article{Dzyaloshinsky1958,
  author    = {Dzyaloshinsky, I.},
  journal   = {Journal of Physics and Chemistry of Solids},
  title     = {{A thermodynamic theory of “weak” ferromagnetism of antiferromagnetics}},
  year      = {1958},
  issn      = {0022-3697},
  month     = jan,
  number    = {4},
  pages     = {241--255},
  volume    = {4},
  doi       = {10.1016/0022-3697(58)90076-3},
  publisher = {Elsevier BV},
}

@Article{Moriya1960,
  author    = {Moriya, Tôru},
  journal   = {Physical Review},
  title     = {{Anisotropic Superexchange Interaction and Weak Ferromagnetism}},
  year      = {1960},
  issn      = {0031-899X},
  month     = oct,
  number    = {1},
  pages     = {91--98},
  volume    = {120},
  doi       = {10.1103/physrev.120.91},
  publisher = {American Physical Society (APS)},
}

@Article{Tanaka2024,
  author    = {Tanaka, Masaaki A. and Yokoyama, Koki and Furuta, Akihiro and Fujii, Kazuki and Mibu, Ko},
  journal   = {Journal of Applied Physics},
  title     = {{Thickness dependence of Morin transition of Ru-doped $\alpha$-Fe$_{2}$O$_{3}$ films detected by spin Hall magnetoresistance measurements}},
  year      = {2024},
  issn      = {1089-7550},
  month     = apr,
  number    = {14},
  pages     = {143901},
  volume    = {135},
  doi       = {10.1063/5.0203488},
  publisher = {AIP Publishing},
}

@Article{Shimomura2015,
  author    = {Shimomura, Naoki and Pati, Satya Prakash and Sato, Yuji and Nozaki, Tomohiro and Shibata, Tatsuo and Mibu, Ko and Sahashi, Masashi},
  journal   = {Journal of Applied Physics},
  title     = {{Morin transition temperature in (0001)-oriented $\alpha$-Fe$_{2}$O$_{3}$ thin film and effect of Ir doping}},
  year      = {2015},
  issn      = {1089-7550},
  month     = apr,
  number    = {17},
  pages     = {17C736},
  volume    = {117},
  doi       = {10.1063/1.4916304},
  publisher = {AIP Publishing},
}

@Article{Gunnink2026,
  author    = {Gunnink, Pieter M. and Sinova, Jairo and Mook, Alexander},
  journal   = {Physical Review Letters},
  title     = {{Surface Acoustic Wave Driven Acoustic Spin Splitter in d -Wave Altermagnetic Thin Films}},
  year      = {2026},
  issn      = {1079-7114},
  month     = mar,
  number    = {11},
  pages     = {116706},
  volume    = {136},
  doi       = {10.1103/y71z-6q6j},
  publisher = {American Physical Society (APS)},
}
\clearpage
	\setcounter{figure}{0}
	\renewcommand{\figurename}{Fig.}
	\renewcommand{\thefigure}{S\arabic{figure}}
	\setcounter{table}{0}
	\renewcommand{\tablename}{Tab.}
	\renewcommand{\thetable}{S\arabic{table}}
	
	\begin{figure*}
		\centering
		\includegraphics{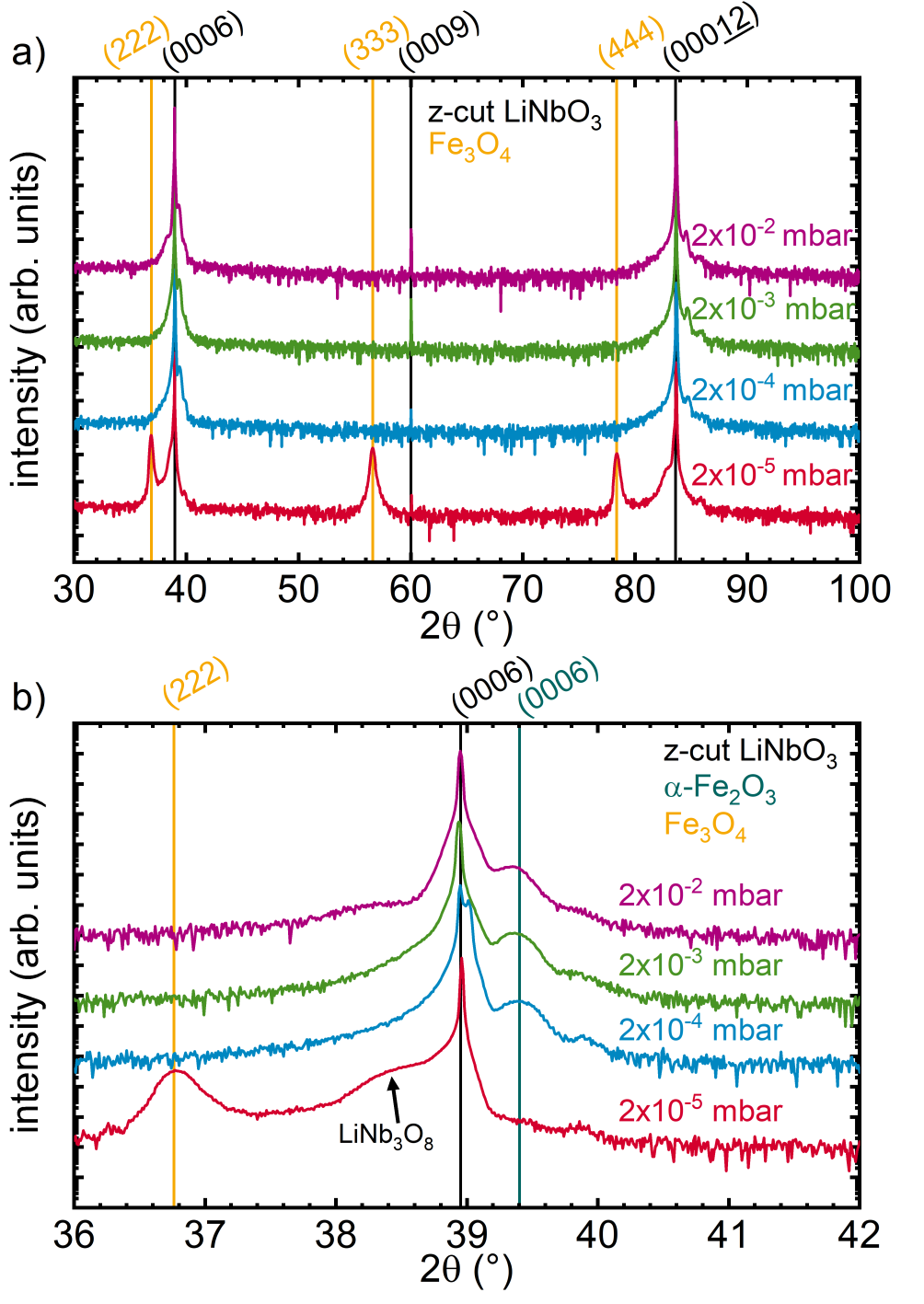}
		\caption{XRD pattern of hematite films deposited at $575\,$°C and a laser fluence of $2.3\,$J$\,$cm$^{-2}$ with different oxygen partial pressures on z-cut lithium niobate a) overview and b) enlargement of the (0006) peak. At an oxygen partial pressure lower than 2\texttimes$10^{-4}\,$mbar Fe$_3$O$_4$ is favored on z-cut lithium niobate. At higher O$_2$ pressures the favored phase is hematite. }
		\label{fig:s1}
	\end{figure*}
	\begin{figure*}[htb]
		\centering
		\includegraphics{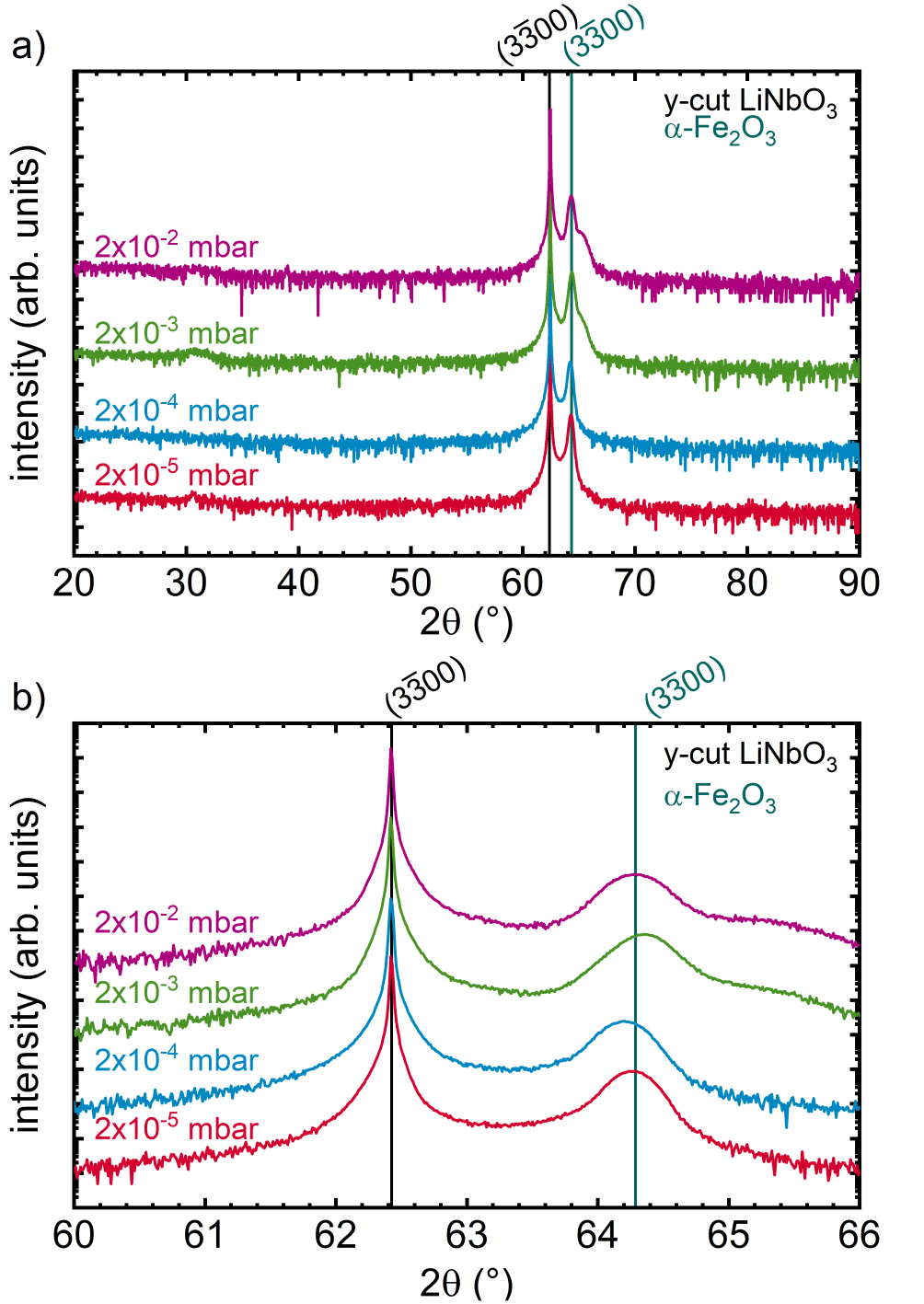}
		\caption{XRD pattern of hematite films deposited at $575\,$°C and a laser fluence of $2.3\,$J$\,$cm$^{-2}$ with different oxygen partial pressures on y-cut lithium niobate a) overview and b) enlargement of the (3$\bar{3}$00) peak. On y-cut LiNbO$_3$, over the whole pressure range, no other iron oxide phase than hematite was observed. }
		\label{fig:s2}
	\end{figure*}
	\begin{table*}[h]
		\centering
		\caption{Film thicknesses of iron oxide films grown on z- and y-cut LiNbO$_3$ at different substrate temperatures with an oxygen partial pressure of 2\texttimes$10^{-4}\,$mbar, and a fluence of $2.3\,$J$\,$cm$^{-2}$, determined by XRR.}
		\begin{tabular}{c|c|c}
			&\multicolumn{2}{c}{Film thickness (nm) of iron oxide films grown on} \\ \hline\hline
			temperature (°C)  & z-cut LiNbO$_3$ &  y-cut LiNbO$_3$\\ \hline
			425 & 59  & 46\\
			475 & fit not possible & 54 \\
			525 & 42 & 39\\
			575 & 56 & fit not possible\\
			625 & 60 (Fe$_3$O$_4$) & 56\\
		\end{tabular}
		\label{tab:s1}
	\end{table*}
	\begin{table*}[h]
		\centering
		\caption{Film thicknesses of iron oxide films grown on z- and y-cut LiNbO$_3$ with different oxygen partial pressures, at a substrate temperature of $575\,$°C, and a fluence of $2.3\,$J$\,$cm$^{-2}$, determined by XRR.}
		\begin{tabular}{c|c|c}
			&\multicolumn{2}{c}{Film thickness (nm) of iron oxide films grown on} \\ \hline\hline
			O$_2$ pressure (mbar) & z-cut LiNbO$_3$ &  y-cut LiNbO$_3$\\ \hline
			2\texttimes$10^{-5}$ & fit not possible (Fe$_3$O$_4$) & fit not possible\\
			2\texttimes$10^{-4}$ &  56 & fit not possible \\
			2\texttimes$10^{-3}$ & 61 & 62\\
			2\texttimes$10^{-2}$ & 52 & 50\\
		\end{tabular}
		\label{tab:s2}
	\end{table*}
	\begin{figure*}[htb]
		\centering
		\includegraphics{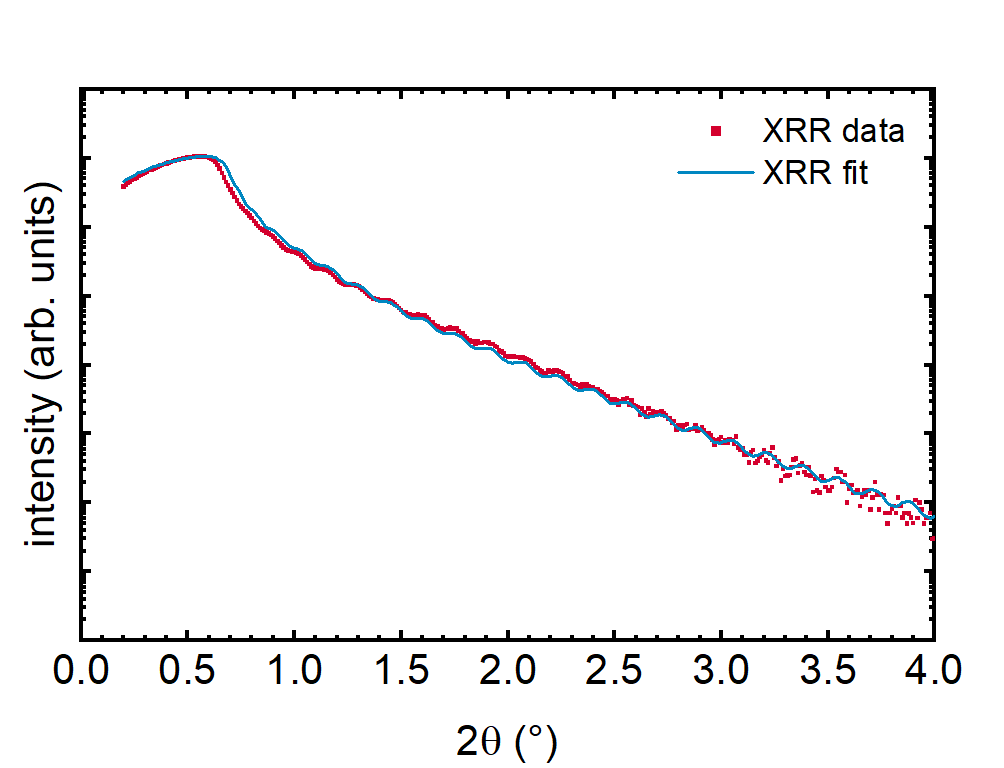}
		\caption{XRR data and fit of a hematite thin film grown on z-cut LiNbO$_3$ at $575\,$°C and an O$_2$ pressure of 2\texttimes$10^{-2}\,$mbar.}
		\label{fig:s3}
	\end{figure*}
	\begin{figure*}[htb]
		\centering
		\includegraphics{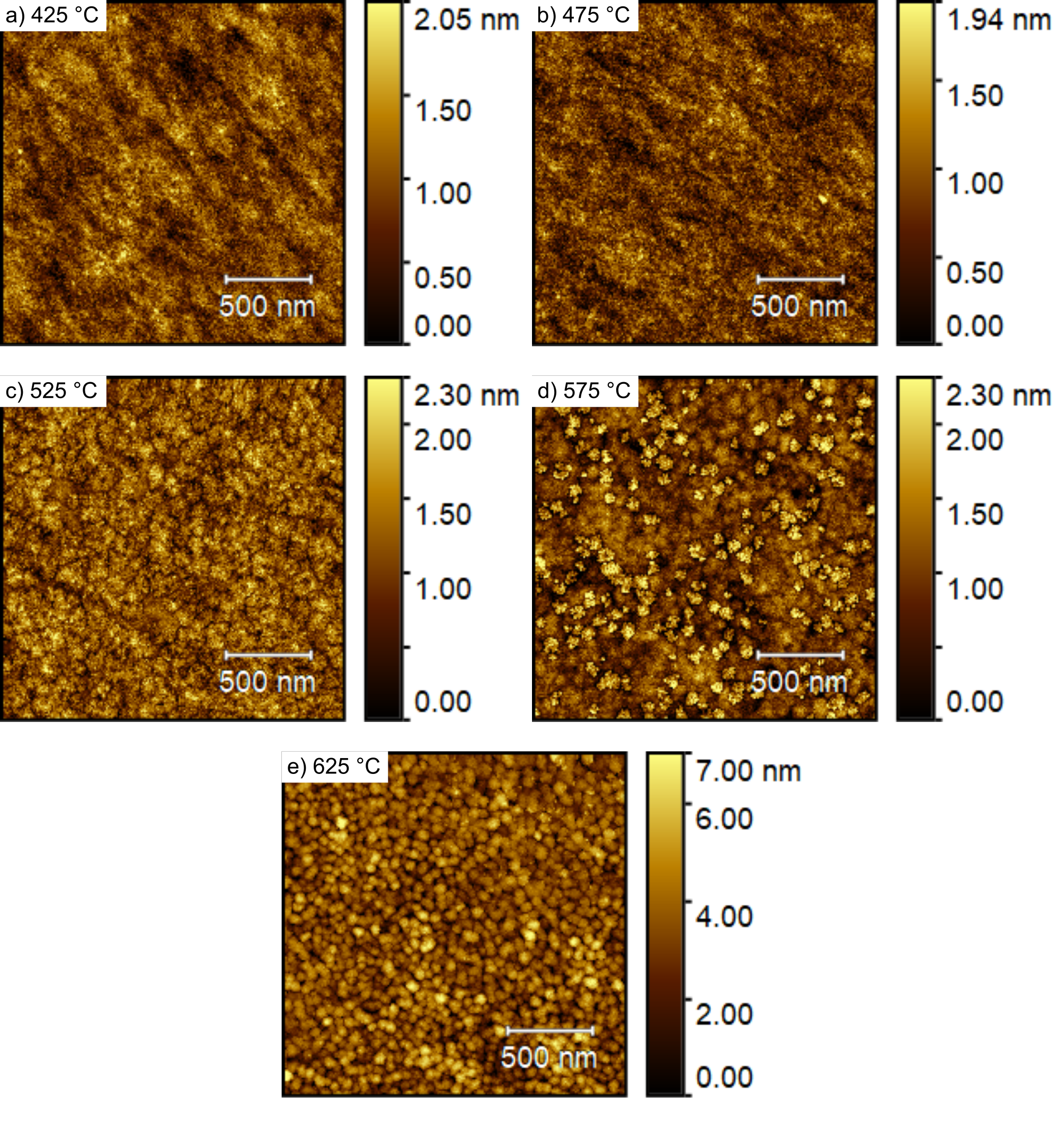}
		\caption{AFM images of iron oxide films grown on z-cut LiNbO$_3$ at 2\texttimes$10^{-4}\,$mbar, $2.3\,$J$\,$cm$^{-2}$, and a substrate temperature of a) $425\,$°C, b) $475\,$°C, c) $525\,$°C, d) $575\,$°C, and e) $625\,$°C.}
		\label{fig:s4}
	\end{figure*}
	\begin{figure*}[htb]
		\centering
		\includegraphics{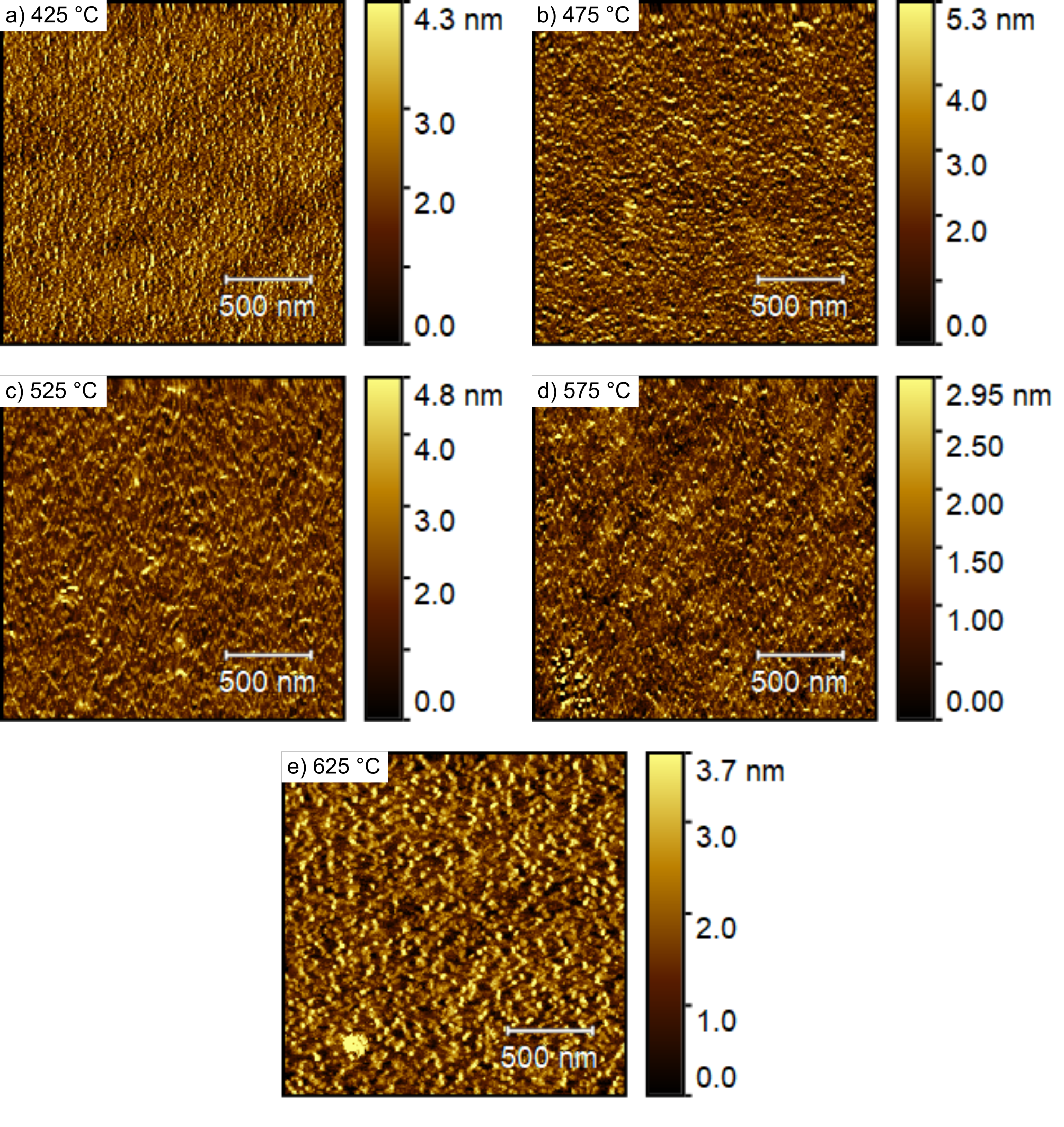}
		\caption{AFM images of iron oxide films grown on y-cut LiNbO$_3$ at 2\texttimes$10^{-4}\,$mbar, $2.3\,$J$\,$cm$^{-2}$, and a substrate temperature of a) $425\,$°C, b) $475\,$°C, c) $525\,$°C, d) $575\,$°C, and e) $625\,$°C.}
		\label{fig:s5}
	\end{figure*}
	
	\begin{figure*}[htb]
		\centering
		\includegraphics{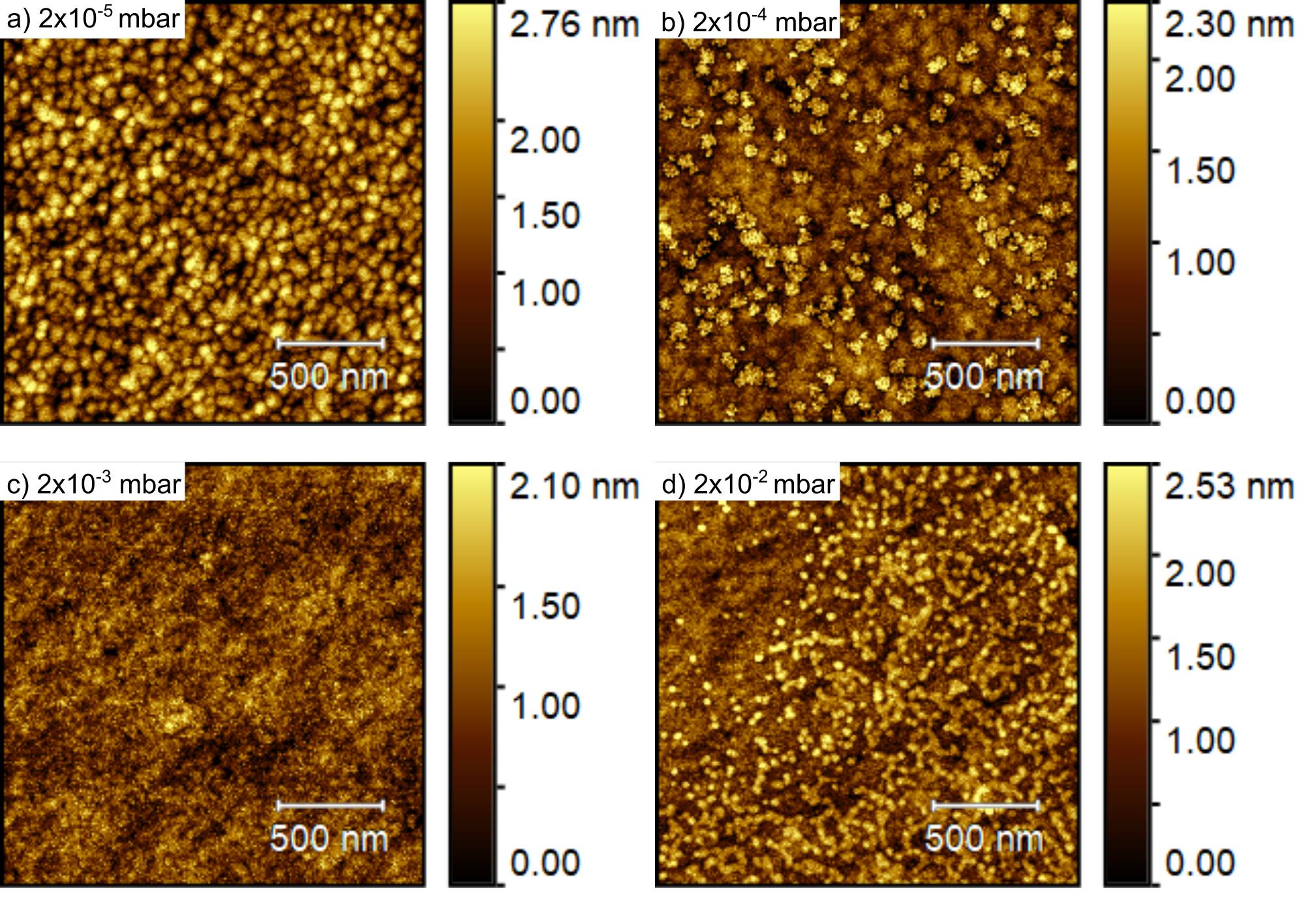}
		\caption{AFM images of iron oxide films grown on z-cut LiNbO$_3$ at $575\,$°C, $2.3\,$J$\,$cm$^{-2}$, and an oxygen partial pressure of a) 2\texttimes$10^{-5}\,$mbar, b) 2\texttimes$10^{-4}\,$mbar, c) 2\texttimes$10^{-3}\,$mbar, and d) 2\texttimes$10^{-3}\,$mbar.}
		\label{fig:s6}
	\end{figure*}
	\begin{figure*}[htb]
		\centering
		\includegraphics{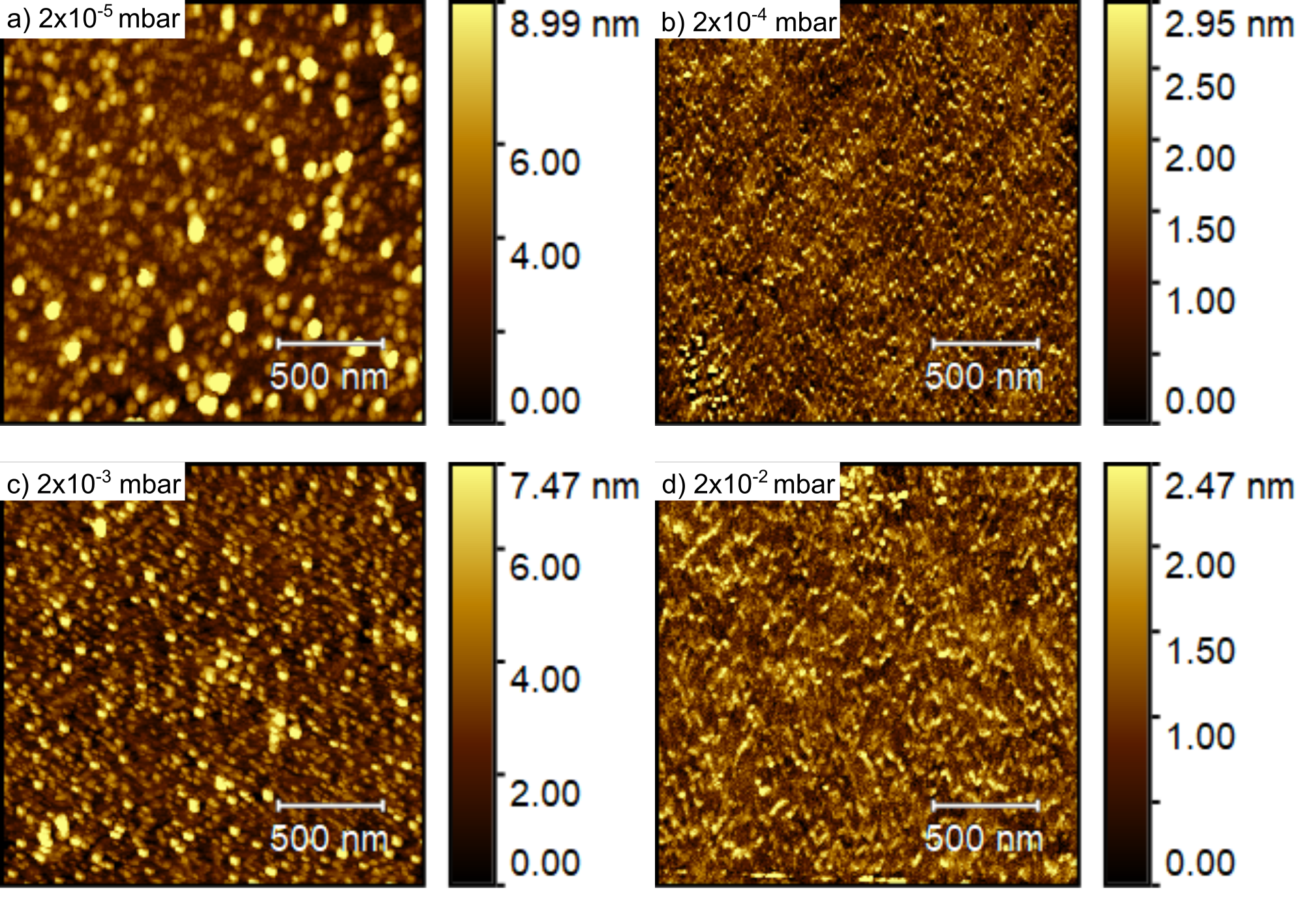}
		\caption{AFM images of iron oxide films grown on y-cut LiNbO$_3$ at $575\,$°C, $2.3\,$J$\,$cm$^{-2}$, and an oxygen partial pressure of a) 2\texttimes$10^{-5}\,$mbar, b) 2\texttimes$10^{-4}\,$mbar, c) 2\texttimes$10^{-3}\,$mbar, and d) 2\texttimes$10^{-3}\,$mbar.}
		\label{fig:s7}
	\end{figure*}
	\newpage
	\begin{table*}[h]
		\centering
		\caption{RMS roughness of iron oxide films grown on z- and y-cut LiNbO$_3$ at different substrate temperatures with an oxygen partial pressure of 2\texttimes$10^{-4}\,$mbar, and a fluence of $2.3\,$J$\,$cm$^{-2}$.}
		\begin{tabular}{c|c|c}
			&\multicolumn{2}{c}{RMS roughness (nm) of iron oxide films grown on} \\ \hline\hline
			temperature (°C)  & z-cut LiNbO$_3$ &  y-cut LiNbO$_3$\\ \hline
			425 & 0.33 & 0.90\\
			475 & 0.32 & 1.08 \\
			525 & 0.40 & 0.80\\
			575 & 0.48 & 0.56\\
			625 & 1.19 & 0.84\\
		\end{tabular}
		\label{tab:s3}
	\end{table*}
	\begin{table*}[h]
		\centering
		\caption{RMS roughness of iron oxide films grown on z- and y-cut LiNbO$_3$ with different oxygen partial pressures, at a substrate temperature of $575\,$°C, and a fluence of $2.3\,$J$\,$cm$^{-2}$.}
		\begin{tabular}{c|c|c}
			&\multicolumn{2}{c}{RMS roughness (nm) of iron oxide films grown on} \\ \hline\hline
			O$_2$ pressure (mbar) & z-cut LiNbO$_3$ &  y-cut LiNbO$_3$\\ \hline
			2\texttimes$10^{-5}$ & 0.64 & 1.80\\
			2\texttimes$10^{-4}$ & 0.40 & 0.80 \\
			2\texttimes$10^{-3}$ & 0.35 & 1.52\\
			2\texttimes$10^{-2}$ & 0.51 & 0.49\\
		\end{tabular}
		\label{tab:s4}
	\end{table*}
	\begin{figure*}[htb]
		\centering
		\includegraphics{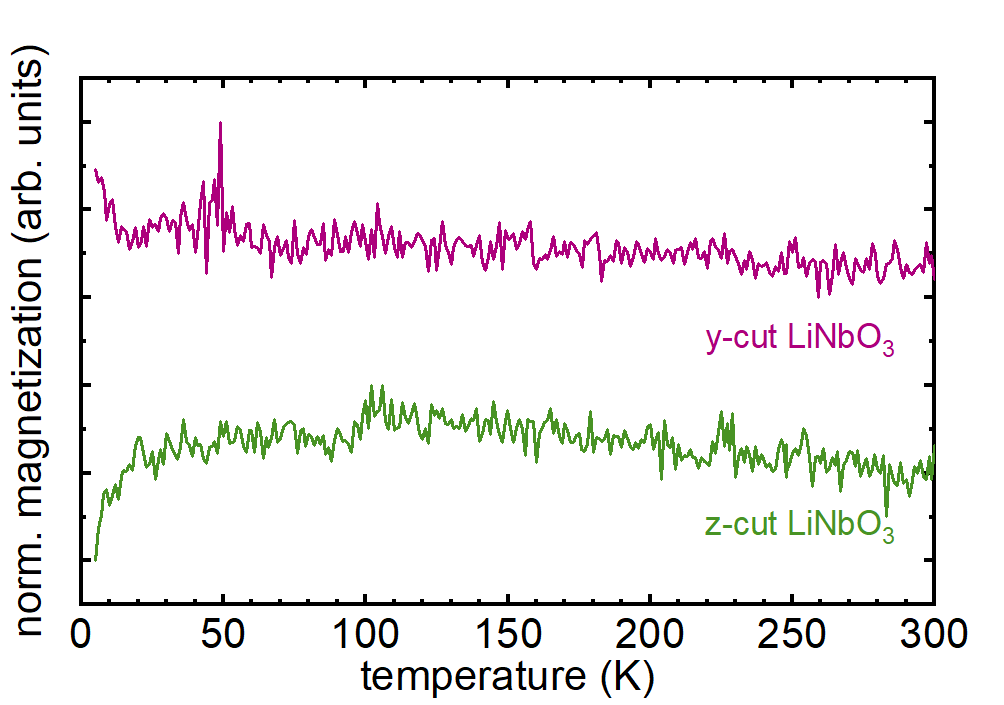}
		\caption{Magnetization versus temperature ($M$ vs $T$) curves of blank z-cut LiNbO$_3$ (green) and y-cut LiNbO$_3$ (purple) substrates. The feature around $50\,$K in the purple curve could originate from internal substrate stress. }
		\label{fig:s8}
	\end{figure*}
\end{document}